\documentclass{article}
\usepackage{hyperref}
\usepackage{graphicx}
\usepackage{amssymb,amsmath}
\usepackage{bbold}
\usepackage{bm}
\usepackage{pgfplots}
\usepackage{tkz-graph}
\usepackage{fixmath}
\usepackage{relsize}
\def\gr{}
\def\rhoo{r}
\def\betaa{b}
\def\INT{_\text{Int}}
\def\AP{{\cal A}_P}
\def\AC{{\cal A}_{\text{Chaos}}}
\def\ALP{{\cal A}_{AP}}

\def\edge{\to}
\newcommand{\bigletter}[1]{\mathlarger{\mathlarger{#1}}}
\newcommand{\bigd}{\bigletter{\vdots}}
\newcommand{\bigO}{\bigletter{N}}

\GraphInit[vstyle = Shade]
\renewcommand*{\VertexBallColor}{green!60}
\renewcommand*{\VertexTextColor}{black}

\tikzset{
 LabelStyle/.style = { rectangle, 
            rounded corners, 
            draw, 
            minimum width = 10em, 
            fill = yellow!50, 
            text = black, },
 VertexStyle/.style = { inner sep=3pt, 
             minimum size = 5pt,
             shape     = \VertexShape,
             ball color   = \VertexBallColor,
             color     = \VertexLineColor,
             inner sep   = \VertexInnerSep,
             outer sep   = 0.5\pgflinewidth,
             text      = \VertexTextColor,
             minimum size  = \VertexSmallMinSize,
             line width   = \VertexLineWidth},
 EdgeStyle/.append style = {->, 
               double=yellow, 
               color=orange}
}

\newcommand{\interior}[1]{%
 {\kern0pt#1}^{\mathrm{o}}%
}
\usepackage{color}

\newcommand{\BF}{\boldmath}

\usepackage{enumitem}

\newcommand{\allblack}{\color{black}{}}

\definecolor{jim}{rgb}{0,0,.5}
\newcommand{\jim}{\color{black} }

\newcommand{\ie}{{\it{i.e.}}}
\newcommand{\eg}{{\it{e.g.}}}

\newcommand{\yz}{\allblack}

\newcommand{\beq}{\begin{linenomath}\begin{equation}} 
\newcommand{\eeq}{\end{equation}\end{linenomath}} 

\def\phi{\varphi}

\def\bZ{\mathbb{Z}}

\def\bR{\mathbb{R}}

\def\cS{{\cal S}}

\usepackage{array}
\newcolumntype{S}{>{\centering\arraybackslash} m{.475\linewidth}}
\newcolumntype{T}{>{\centering\arraybackslash} m{10.5cm}}
\newcolumntype{U}{>{\centering\arraybackslash} m{1.5cm}}
\title{
Infinite towers in the graph of a dynamical system
}
\author{Roberto De Leo\thanks{Department of Mathematics, Howard University, Washington DC 20059, roberto.deleo@howard.edu}\ \ and James A. Yorke\thanks{Institute for Physical Science and Technology and the Departments of Mathematics and Physics, University of Maryland College Park, MD 20742, yorke@umd.edu}}
\usepackage{lineno}

\begin{document}
\maketitle

\begin{abstract}
 Chaotic attractors, chaotic saddles and periodic orbits are examples of chain-recurrent sets. 
  Using arbitrary small controls, a trajectory starting from any point in a chain-recurrent set can be steered to any other in that set. 
 The qualitative behavior of a dynamical system can be encapsulated in a graph. Its nodes are chain-recurrent sets. There is an edge from node $A$ to node $B$ if, using arbitrary small controls, a trajectory starting from any point of $A$ can be steered to any point of $B$. 
 We discuss physical systems that have infinitely many disjoint coexisting nodes. 
 Such infinite collections can occur for many carefully chosen parameter values. The logistic map is such a system, as we show in a rigorous companion paper. 
 To illustrate these very common phenomena, we compare the Lorenz system and the logistic map and 
 we show how extremely similar their {\jim graph} bifurcation diagrams are in some parameter ranges. {\jim
 Typically, bifurcation diagrams show how attractors change as a parameter is varied. 
 We call ours ``{\bf graph bifurcation diagrams}'' to reflect that not only attractors but also unstable periodic orbits and chaotic saddles can be shown. 
 Only the most prominent ones can be shown.
 We argue that, as a parameter is varied in the Lorenz system, there are uncountably many parameter values for which there are infinitely many nodes, and infinitely many of the nodes  $N_1,N_2,N_3,\ldots,N_\infty$ can be selected so that the graph has an edge from each node to every node with a node with a higher number. The final node $N_\infty$ is an attractor}.
\end{abstract}
\newpage
\section{Introduction and definitions}

In 1970's, 
Charles Conley introduced the idea of describing the qualitative behavior of a dynamical system by the type of graph that we describe below. 
In~\cite{DLY20} we show that the graph of the logistic map $\mu x(1-x)$ is surprisingly complicated for certain values of $\mu$. Here we argue that the most complicated logistic map graphs appear within the graphs of much more general and complicated systems. 
To illustrate this fact, we compare the logistic map with the Lorenz system {\jim using non-rigorous numerical investigations.}


We alert the reader that there is a similarity between some of the pictures in this paper and in~\cite{DLY20}.

It might seem to the reader that the Lorenz system and the logistic map appear to be completely unrelated. 
That is why we have selected the Lorenz system, when we could have chosen any of a wide variety of physical systems. On the other hand, we have chosen the logistic map because of the rich rigorous literature that is available on it. 

Bifurcation diagrams for the logistic map typically show how the attractor changes as a parameter changes.
In addition to an attractor, the logistic map has several other disjoint invariant sets, and there are  parameter values for which there are infinitely many of them. The invariant sets we speak of are ``chain-recurrent'', as we describe below.

{\bf An example of dynamical system with a simple graph.}
Consider the  
 map $z\mapsto z^2$ 
 on the complex plane, to which we add the point at $ \infty$. The plane plus $\infty$ should be thought of as a topological sphere. For many important cases, we can ``compactify'' a space by adding a point at $\infty$ and often, as for this map, $\infty$ is a fixed point and the map is still continuous.

We can use this map as an example of how to represent a dynamical system by a graph.
This map has three invariant sets that will be nodes of the graph. 
Both $\{0\}$ and $\{\infty\}$ are attractors and are nodes, and the third node is the unit circle, a repelling chaotic invariant set. 
Notice that not all invariant sets are nodes.
Explaining what a node is will take some care. 
Even for such a simple map, the dynamics {\em within} a node can be quite complicated. 
For instance, in the $z^2$ example, the dynamics on the unit circle $z=e^{i\phi}$, $\phi\in[0,2\pi]$, is given by the {\em doubling map} $\phi\mapsto2\phi$ (also known as {\em shift} or {\em Bernoulli} map). This map is one of the best known examples of a chaotic map. 
Notice that there are infinitely many periodic orbits on the unit circle but none of them is a node.
The set of nodes of a general dynamical systems can be quite a bit more complicated than the set of three nodes in this case. 

This paper is about a type of control theory. For each point $p$, it identifies the downstream point $q$ such that either the trajectory from $p$ goes to $q$ or an arbitrarily small amount of control can be added such that the controlled trajectory goes from $p$ to $q$. 
We now extend the stream analogy.
If $p$ is downstream from $q$ and $q$ is downstream from $p$, then we say $p$ and $q$ are in the same pond.
A {\bf node} is a pond. In other words, a node $N$ is the set of points so that, if $p$ is in $N$, then $q$ is in $N$ if and only if $p$ and $q$ are in the same pond.
A trajectory starting form any point in the node can be forced to stay in the node by using arbitrarily small perturbations that we call controls.
We make this precise as follows.

\begin{figure}
 \centering
 \includegraphics[width=6.5cm]{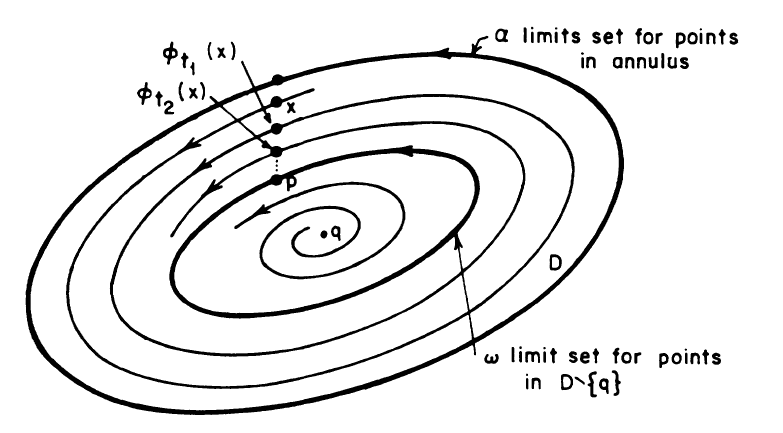}
 \includegraphics[width=4.5cm]{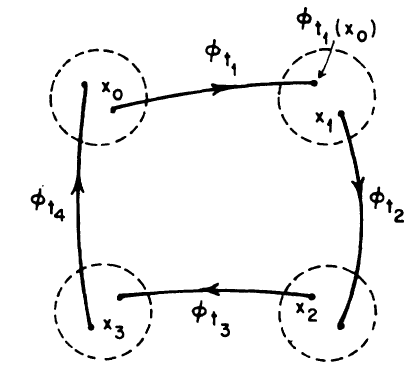}
 \caption{{\bf Examples of chain-recurrence. (LEFT)} Example of nodes in a continuous dynamical system. The set $D$ is the disc bounded by the outer periodic orbit. 
Three nodes are visible in the picture: the
outer periodic orbit $N_1$, the inner periodic orbit $N_2$ and a fixed point $N_3$. 
 The edges of this \gr graph go from the repellors $N_3$ and $N_1$ to the attractor $N_2$.
 {\bf (RIGHT)} An example of an $\varepsilon$-chain from $x_0$ and back to itself. The dashed circles represent circles of radius $\varepsilon$.}
 \label{fig:cr}
\end{figure}

{\bf Chain-recurrence.} By a dynamical system $\Phi$, we mean a 1-parameter family of continuous maps $\Phi^t$ from a space $X$ into itself. 
Write $dist(x,y)$ for the distance between $x$ and $y$. 
The time parameter $t$ can be either continuous or discrete. 
Given two points $p,q$, with $p\ne q$, in $X$ and $\varepsilon>0$, we say that there is a {\bf\BF$\varepsilon$-chain} from $p$ to $q$ (see Fig.~\ref{fig:cr}) if there is a finite sequence of points $p=x_0,x_1,\dots,x_n=q$ on $X$ such that, for $i=0,\dots,n-1$,
\beq
 \mbox{dist}(\Phi^{1}(x_i),x_{i+i})\le\varepsilon.
 \label{chain}
\eeq
To our knowledge, $\varepsilon$-chains were introduced in literature by R.~Bowen in 1975~\cite{Bow75}.

We say that {\bf \BF $q$ is downstream from $p$} if, for every $\varepsilon> 0$, there is a $\varepsilon-$chain from $p$ to $q$; equivalently, we say that {\bf \BF $p$ is upstream from $q$}. 
We write {\BF$p\sim q$} if $p$ is upstream and downstream from $q$, and we say that {\bf \BF$p$ is chain recurrent} if $p\sim p$. 
We let {\BF${\cal R}_\Phi$} denote the {\bf chain-recurrent set}, \ie\ the set of all chain recurrent points of $\Phi$. 
Chain recurrence was introduced by C.~Conley in his celebrated monograph in 1978~\cite{Con78} and it is a central concept for this article.

{\bf Examples of chain-recurrent points}. Points on a periodic orbit are chain-recurrent and, if $p$ and $q$ are on the same periodic orbit, then $p\sim q$.

Chaotic sets are defined in various ways but a usual requirement is that there is a trajectory that comes arbitrarily close to every point infinitely often. So, if $p$ and $q$ are in a chaotic set, a tiny perturbation of $p$ will land on the dense trajectory and, when it comes sufficiently close to $q$, a second tiny perturbation will push it onto $q$. Hence, $p\sim q$. 

Consider now a dynamical system on a vector space and suppose that all trajectories converge to 0 as time goes to infinity. Then 0 is the only chain-recurrent point.

{\bf Subtle control of dynamical systems.} The idea of $\varepsilon-$chains in (\ref{chain}) can be rephrased as the following question in control theory. Assume $X$ is in a linear space.
Given two points $p,q,p\ne q,$ in $X$, does there exist for each 
$\varepsilon>0$ a finite sequence of $u_i$ such that
$|u_i|\le\varepsilon$ for a sequence of $i$'s
and a controlled trajectory
\beq
x_{i+i}=\Phi^{1}(x_i)+u_i \mbox{ where }p=x_0,x_n=q.
\label{control}
\eeq
If $p\sim q$, then there are such controls and it is possible to create control $u_i$ that allow us to steer a trajectory from $p$ to $q$ and back to $p$. Furthermore, $\max|u_i|$ can be made as small as desired, \ie, less than any specified positive number.

\allblack

\begin{figure}
 \centering
 \includegraphics[width=5.5cm]{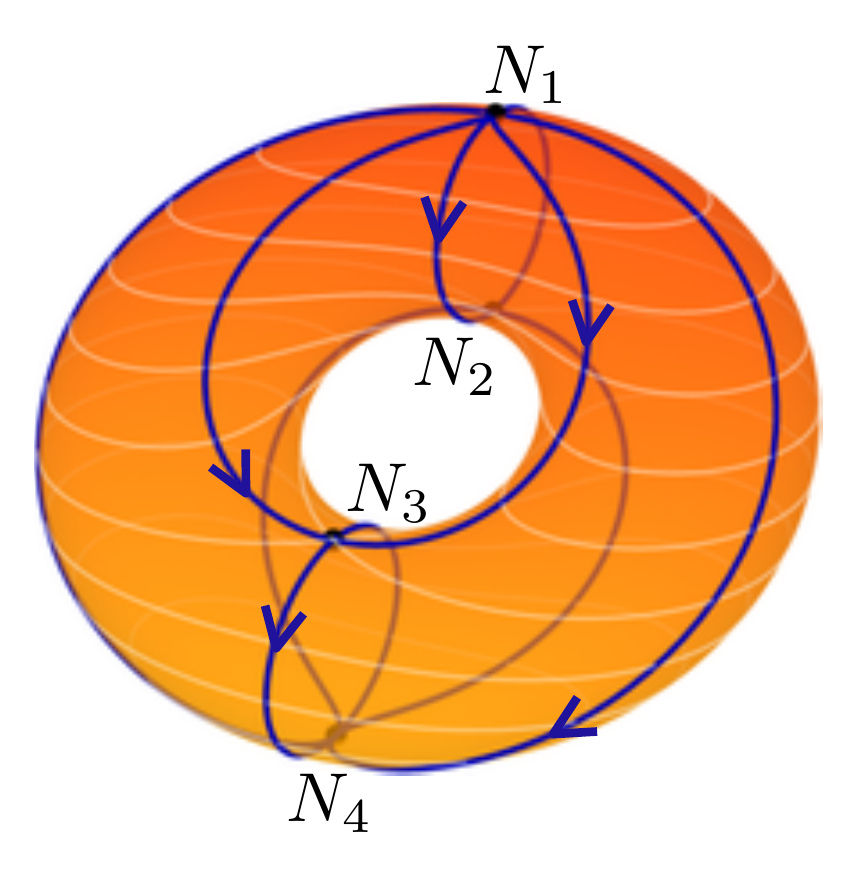}\hskip2.cm 
 \begin{tikzpicture}[scale=.3]
 \SetGraphUnit{5}
 \Vertex[L=$\bigO_1$]{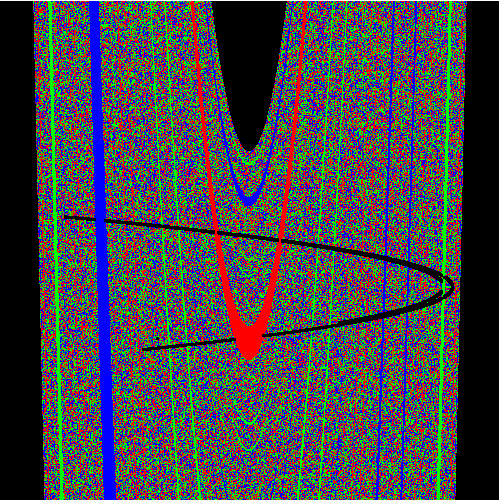}
 \SO[L=$\bigO_2$](A){B}
 \SO[L=$\bigO_3$](B){C}
 \SO[L=$\bigO_4$](C){D}
 \tikzset{EdgeStyle/.append style = {bend left=45}}
 \Edge(A)(B)
 \Edge(A)(C)
 \Edge(A)(D)
 \tikzset{EdgeStyle/.append style = {bend right=45}}
 \Edge(B)(C)
 \Edge(B)(D)
 \tikzset{EdgeStyle/.append style = {bend right=0}}
 \Edge(C)(D)
\end{tikzpicture}
 \caption{{\bf An example of \gr graph.} {\bf (LEFT)} Dynamics induced on the 2-torus by the gradient vector field of the height function. In this case the Lyapunov function is the height function itself, some level set of which is shaded in white. In blue are shown the heteroclinic trajectories joining the critical point (which are exactly the invariant sets of this dynamical system). {\bf (RIGHT)} The \gr graph of the dynamical system on the left. In this case it is a 4-levels tower.
 }
 \label{fig:pp}
\end{figure}
  
{\bf A trajectory of a dynamical system.}
Here we restrict attention to discrete time dynamical systems. For a map $\Phi$,
we will say that the sequence $p_n$ is a {\bf trajectory} if
 $p_n$ is defined for all $n\in\bZ$,
where $\bZ$ is the set of all integers, $n=0,\pm1,\pm2,\ldots$, and 
$p_{n+1}= \Phi(p_n)$ for all $n\in \bZ$.

For some maps, the inverse is not unique. For the map $z\mapsto z^2$, each point other than $0$ has two inverses. 
Hence there will be infinitely many trajectories through a given $p_0 \ne 0.$ Two different trajectories through $p_0$ will have the same forward limit set but might have different backward limit sets.
\begin{figure}
 \centering
\tikzset{
 LabelStyle/.style = { rectangle, rounded corners, draw, minimum width = 2em, text = black, font = \bfseries },
 EdgeStyle/.append style = {->, bend right=62,double}
}
\scalebox{1}{
\begin{tikzpicture}[scale=.3]
 \SetGraphUnit{5}
 \Vertex[L=$\bigO_1$]{A}
 \Vertex[L=$\bigO_{m}$,a=-90,d=5.1cm]{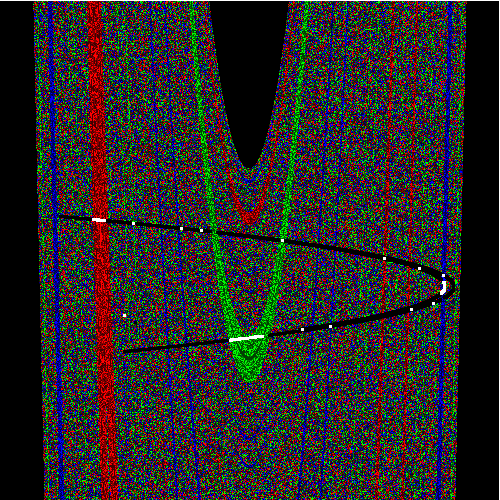}
 \Vertex[L=$\bigO_{n}$,a=-90,d=11.cm]{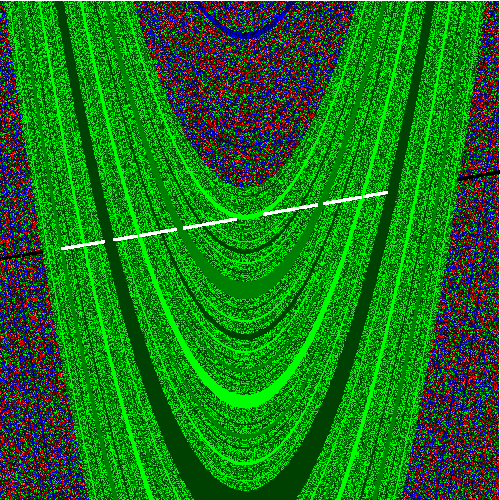}
 \Vertex[L=$\bigO_\infty$,a=-90,d=16.5cm]{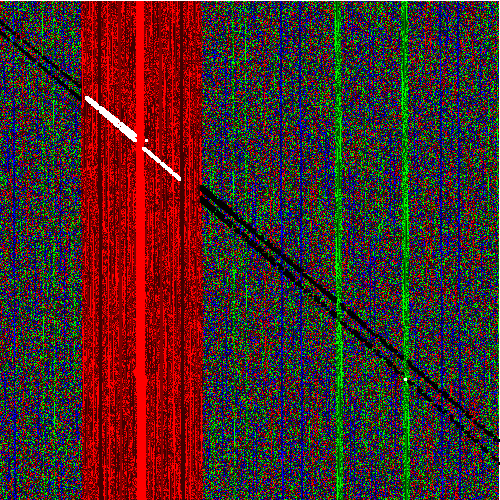}
 \Edge[local=true,label=$\bigd$,color=white,labelstyle={fill=white},style={bend left=0}](A)(B)
 \Edge[local=true,label=$\bigd$,color=white,labelstyle={fill=white},style={bend left=0}](B)(C)
 \Edge[local=true,label=$\bigd$,color=white,labelstyle={fill=white},style={bend left=0}](C)(D)
 \Edge(A)(B)
 \Edge(A)(C)
 \Edge(A)(D)
 \tikzset{EdgeStyle/.append style = {bend left=45}}
 \Edge(B)(C)
 \Edge(B)(D)
 \Edge(C)(D)
\end{tikzpicture}
}
 \hskip2.cm 
\scalebox{.9}{
\begin{tikzpicture}[scale=.275]
 \presetkeys[GR]{vertex}{LabelOut = true}{}  
 \SetGraphUnit{5.75}
 \tikzset{EdgeStyle/.append style = {->, bend left=0,double}}
 \Vertex[L=$\bigO_1$]{O}
 \Vertex[L=$\bigO^+_2$,a=-135,d=5.3cm,Lpos=180]{A1}
 \Vertex[L=$\bigO^-_2$,a=-45,d=5.3cm]{B1}
 \SO[L=$\bigO^+_3$,Lpos=180](A1){A2}
 \SO[L=$\bigO^-_3$](B1){B2}
 \SO[L=$\bigO^+_{n-1}$,Lpos=180](A2){A3}
 \SO[L=$\bigO^-_{n-1}$](B2){B3}
 \SO[L=$\bigO^+_n$,Lpos=180](A3){A4}
 \SO[L=$\bigO^-_n$](B3){B4}
 \Edge(O)(A1)
 \Edge(O)(B1)
 \Edge(A1)(A2)
 \Edge(B1)(B2)
 \Edge(A1)(B2)
 \Edge(B1)(A2)
 \Edge(A3)(A4)
 \Edge(B3)(B4)
 \Edge(A3)(B4)
 \Edge(B3)(A4)
 \Edge[local=true,label=$\bigd$,color=white,labelstyle={fill=white},style={bend left=0}](A2)(A3)
 \Edge[local=true,label=$\bigd$,color=white,labelstyle={fill=white},style={bend left=0}](B2)(B3)
\end{tikzpicture}
}
 \caption{{\bf Examples of \gr graphs. (LEFT)} An infinite tower \gr graph. {\bf (RIGHT)} The \gr graph of the semiflow of the Chafee-Infante PDE (see Sec.~\ref{sec:inf}). 
 }
 \label{fig:co}
\end{figure}

{\bf Assumptions on the phase space.}
In this paper, aside from our infinite dimensional examples, we examine continuous dynamical systems on a compact set $X$.
In the above example, we have added a point at infinity to make the set compact. In this paper we use the following definition. A set $X$ is compact if for each sequence of points $x_n (n = 1,2,\dots,\infty)$, there is a subsequence $x_{n_j} ~(j = 1,2,\dots,\infty)$ that converges to some point $p$. Considering all convergent subsequences, the set of limit points $p$ is the limit set of 
$x_n$.

{\bf Where are the limit sets.}
For any point $x$, its forward limit set $\omega(x)$ is the set of its limit points, namely those points that are the limit a subsequence of points belonging to the forward orbit of $x$. 
Its trajectory might diverge, \ie, its limit set is empty. Then we can say it converges to the node $\infty.$
Otherwise, its limit set 
must be a subset of a single node.
For example picture a situation where a trajectory in the plane lies between two invariant lines and it spirals outward toward those lines. 
Then the node includes all the points on those two lines.
If that node $\Omega$ is a compact set, then the distance of $\Phi^t(x)$ from $\Omega$ goes to $0$ as $t\to\infty.$

{\bf Attractors.} We call a node $N$ an {\bf attractor}, also sometimes called a {\bf Milnor attractor}, if its basin of attraction, \ie, the set of points $x$ such that $\omega(x)$ is contained in $N$, has positive measure~\cite{Mil85}. 
A non-trivial example of a Milnor attractor occurs at the Feigenbaum parameter value. 

{\bf The graph of a dynamical system and Lyapunov functions.} 
Conley realized that chain-recurrence could be used to define a graph of a dynamical system~\cite{Con72,Con78}.
His investigations concerned dynamical systems that come from ordinary differential equations on compact spaces.
Over the years his results have been extended to several other settings, in particular: continuous maps~\cite{Nor95b}, semi-flows~\cite{Ryb87,HSZ01,Pat07}, non-compact~\cite{Hur91,Pat07} and even infinite-dimensional spaces~\cite{Ryb87,MP88,CP95,HMW06} (here and  throughout this article, we sort multiple citations in the order of their year of publication).

The main contribution of Conley is the discovery that the dynamics outside of the nodes is always {\bf gradient-like}, namely there is a continuous function $L:X\to\bR$ such that:
\begin{enumerate}
  \item $L$ is constant on each node; 
  \item $L$ assumes different values on different nodes;
  \item $L(\Phi^tx)< L(x)$ for all $t>0$ and when $x$ not in a node~\cite{MM02}.
\end{enumerate}
In particular, nodes are equilibria for $L$. 
Note also that properties 1, 2 and 3 make $L$ a {\bf Lyapunov function} (\eg~see~\cite{WY73,Nor95b}).

The graph of a dynamical system consists of nodes and edges between the nodes. 
The forward and backward limit sets of a trajectory are each contained inside a single node. That limit set can also be the entire node. 
There is an {\bf edge} from node $N_1$ to node $N_2$ if and only if there is a trajectory whose backward limit set is in $N_1$ and its forward limit set is in $N_2$ (\eg, see Fig.~\ref{fig:pp} (right) and Fig.~\ref{fig:co}). 
That edge can be denoted by {\BF $N_1\edge N_2$}, which reads that $N_1$ is above $N_2$.
In particular, $N_1\edge N_2$ implies that $L(N_1)>L(N_2)$, so that it is impossible that also $N_2\edge N_1$.

Each node $N$ has a closest point to the critical point $c=1/2$.
Let $\rho(N)$ denote the distance between $c$ and that closest point.
We show in~\cite{DLY20} that $N_1\edge N_2$ is equivalent to saying $\rho(N_1)>\rho(N_2)$.

%
\begin{figure}
 \centering
 \includegraphics[width=12cm]{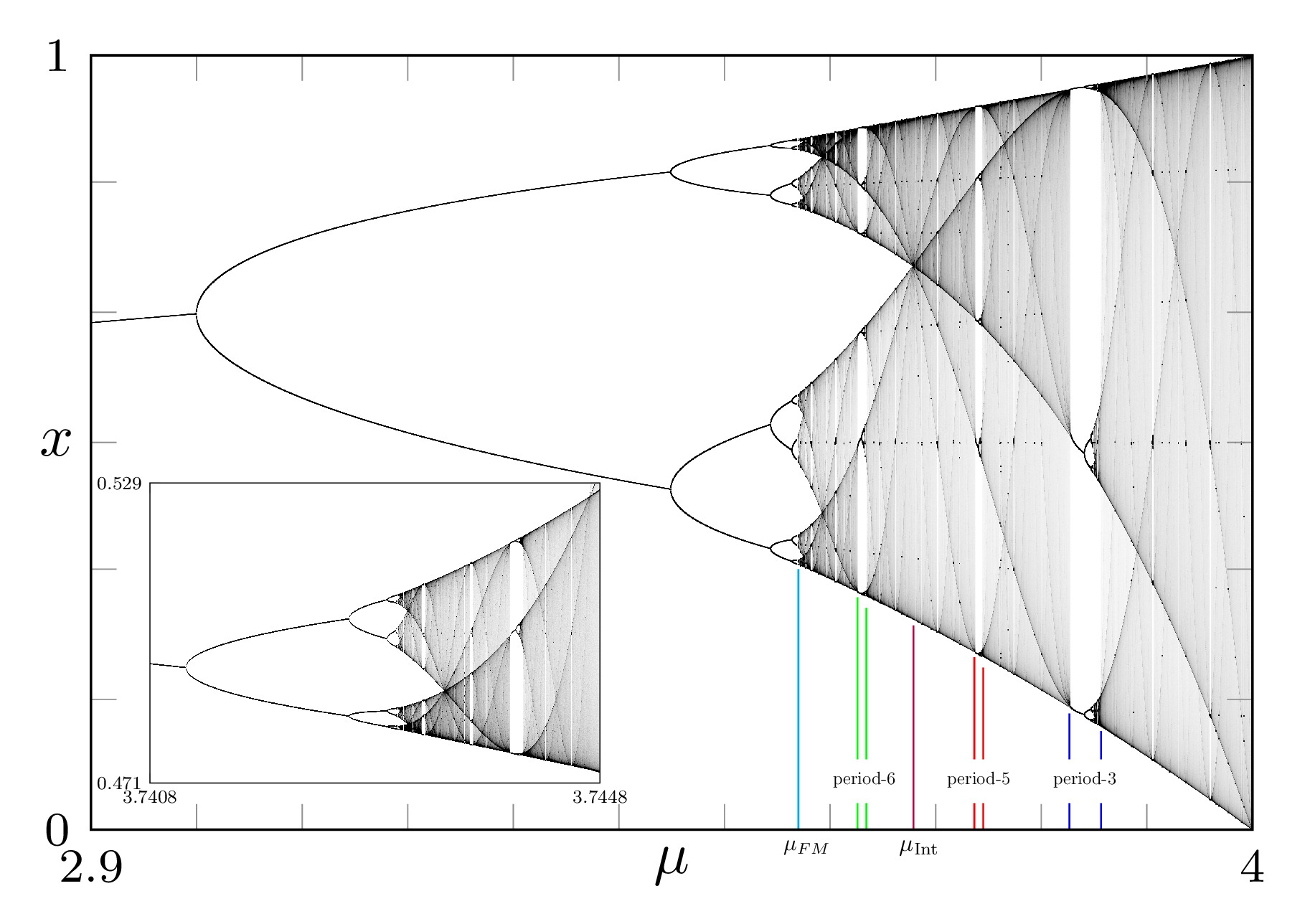}
 \caption{{\bf The Bifurcation Diagram of the Logistic Map.} To the left of the Feigenbaum-Myrberg parameter value $\mu_{FM}\simeq3.56994567$, we see the well-known period-doubling cascade. To its right, we see lots of chaos but also many windows, \ie, intervals in parameter space that begin with a periodic attractor which evolves through period-doubling into small intervals of chaos. This picture is created by plotting trajectories. More frequently visited regions are darker. Points on attracting periodic orbits of period less than 26 are indicated by black dots. Notice, in particular, that many of these points are near where $x=0.5$. In colors are highlighted, besides $\mu_{FM}$, the largest period-6 window, the intersection parameter value $\mu\INT=1.5[ 1 + (19-3\sqrt{33})^{1/3} + (19+3\sqrt{33})^{1/3} ]\simeq3.67857$, the largest period-5 window and the largest period-3 window. Notice that many high-density lines intersect at $(\mu\INT,x\INT)$ (see figure). Each of the high-density lines is the image of the $x=0.5$ line under $\ell^n_\mu$ for some $n$. In the bottom right box it is shown a detail of the cascade about $x=0.5$ inside the period-5 window.
 }
 \label{fig:lmbd}
\end{figure} 

{\jim
{\bf Any edge in a graph can be thought of as a set of points.}
The {\bf unstable set} of a node $N$ is the set of points $X$ such that for each $\varepsilon >0$, there is an $\varepsilon$-chain from a point in $N$ to $X$.
The {\bf stable set} of a node $N$ is the set of points $X$ such that for each $\varepsilon >0$, there is an $\varepsilon$-chain from $X$ to a point in $N$.
The edge from node $N_1$ to node $N_2$ can be identified with the points X that are on both the unstable set of $N_1$ and the stable set of $N_2$.
If we have 3 nodes $N_1\to N_2\to N_3$, the set $N_1\to  N_3$ includes
$N_1\to N_2$ and $N_2\to N_3$ and possibly other points.

{\bf An alternative way to define a graph.} In this paper we follow Conley's definition of a graph, where nodes are defined in terms of $\varepsilon$-chains while edges are defined in terms of stable and unstable sets.
Any interested reader could choose instead to define edges in terms of $\varepsilon$-chains, and that might make proofs easier. 
If one defines edges in terms of $\varepsilon$-chains, our results stated here still hold because what was an edge is still and edge, though additional edges can be created in other systems. 
Then, if $N_1\to N_2$ and $N_2\to N_3$, with the chain-recurrent definition, one automatically has $N_1\to  N_3$.
}

\yz
{\bf Graphs in 1-D.} The classification of more complex nodes was an important milestone even in the setting of 1-dimensional dynamics (a list of specific references is given in Sec.~\ref{sec:lm}). 
In this last case, though, it seems that the dynamical system community put the emphasis in the classification of the nodes and somehow overlooked the description of the rest of the dynamics,
that is, which pairs of nodes $N_1,N_2$ have an edge {$N_1\edge N_2$}.

\begin{figure}
 \centering
 \includegraphics[width=12cm]{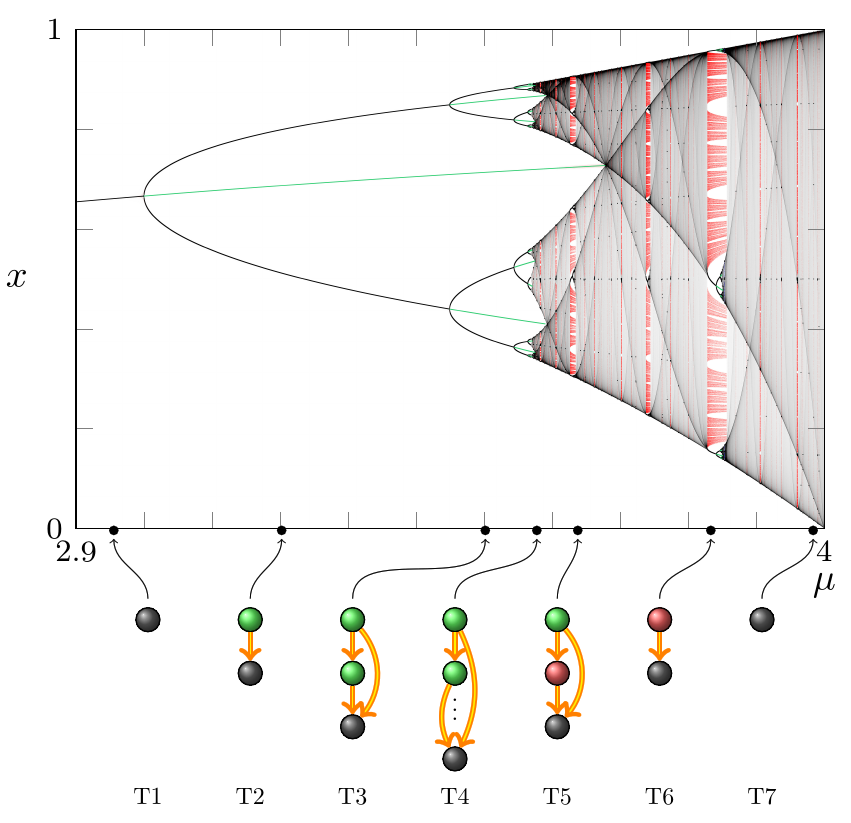}
  \caption{{\bf Bifurcation diagram and sample graphs of the logistic map.} 
  This picture shows the bifurcation diagram of the logistic map in the range of parameter values $[2.9,4]$.
  For each value of $\mu$, the attracting set is painted in shades of gray, depending on the density of the attractor, repelling periodic orbits in green and repelling Cantor sets in red. Below the $\mu$ axis we show seven samples of the graphs illustrating some of the possible variability. In these graphs, each colored disk is a node. Each black disk represents an attractor, each green disk represents a repelling periodic orbit and red represents a chaotic Cantor set repellor. For simplicity we always omit the top node, which is the point 0. Graph T4 represents the infinite tower at the first Feigenbaum point. It has infinitely many unstable periodic orbit nodes.
  }
  \label{fig:full}
\end{figure}


{\bf Towers.} We call a {\bf tower} a finite or infinite sequence of nodes $N_i$ such that: 
\begin{enumerate}
    \item there is a first node, denoted by $N_0$;
    \item there is a final node, which is the unique attractor; all other nodes are unstable;
    \item for any two nodes $N_i$ and $N_j$, with $j>i$, we have that {$N_i\edge N_j$}.
\end{enumerate}
In particular, for each node $N_i$, where $i>0$, there is a previous node $N_{i-1}$ and, unless $N_i$ is the attractor, a next node $N_{i+1}$.

Our main result in~\cite{DLY20} is the following:

\smallskip
{\bf Logistic Tower Theorem.} 
{\jim For each parameter value $\mu\in(1,4]$, the graph of the logistic map is a tower.}
\smallskip

\begin{figure}
 \centering
 \includegraphics[width=11cm]{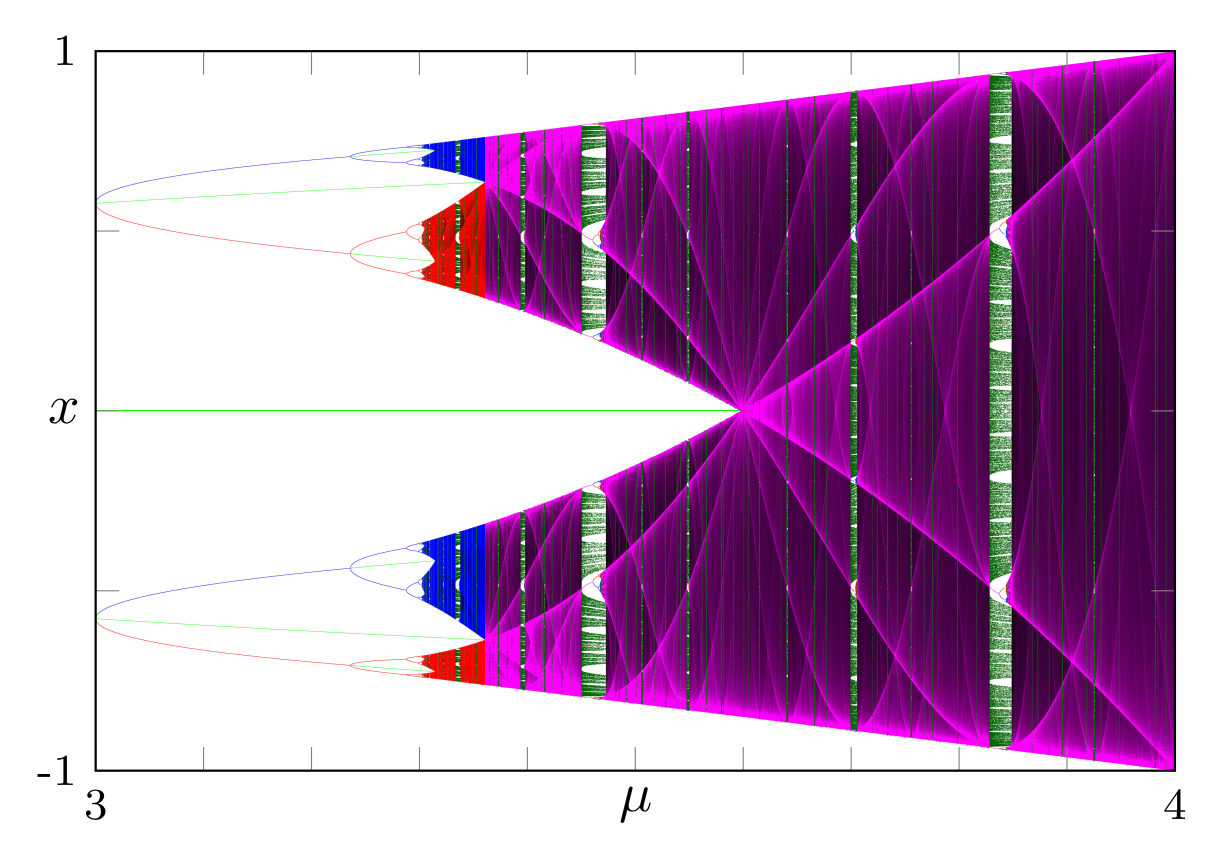}
 \caption{{\bf\jim Graph bifurcation diagram of a one-dimensional map with two critical points.}
 \jim This figure is the graph bifurcation diagram of the function $b_\mu(x)=x-\mu x(1-x^2)$, that maps the interval $[-1,1]$ into itself for every $0\leq\mu\leq4$. 
 For each value of $\mu$, $b_\mu$ has three fixed points, namely $x=0,\pm1$, and either one or two attractors. When there are two attractors, we paint one in blue and one in red. When there is a single attractor, we paint it in purple. The light green points belong to unstable periodic orbits, the dark green ones to chaotic unstable Cantor sets. 
 Lighter purple implies higher trajectory density than darker purple.
 The light purple lines correspond to infinite density.
}
 \label{fig:S-bimodal}
\end{figure} 

For specific parameters, the logistic map has infinitely many nodes.
In this case, we refer to the tower as an {\bf infinite tower}. 
{\jim 
We believe infinite towers are common in higher-dimensional systems, but we would expect that the infinite towers are subsets of more complex graphs.
}

\begin{figure}
 \centering
 \includegraphics[width=13cm]{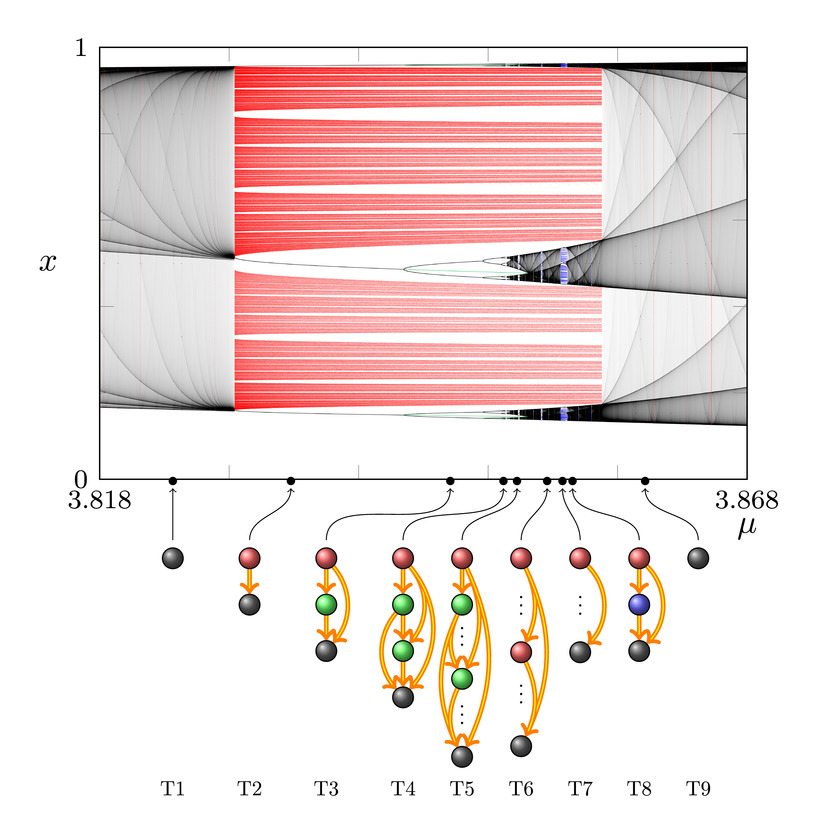}
 \caption{{\bf Towers of nodes shown below the period-3 window of the logistic map graph bifurcation diagram.}
    This figure is a blow-up from Fig.~\ref{fig:full}, and uses the color-coding from that figure. Graph T8 has two levels of nodes that are Cantor sets repellors and the second is painted in blue.
In the bifurcation diagram, the chain recurrent sets have the same coloring as their nodes. 
Graph T5 represents the infinite tower at the first Feigenbaum point of the main cascade of the period-3 window. 
}
 \label{fig:p3}
\end{figure}

\begin{figure}
 \centering
 \includegraphics[width=12.5cm]{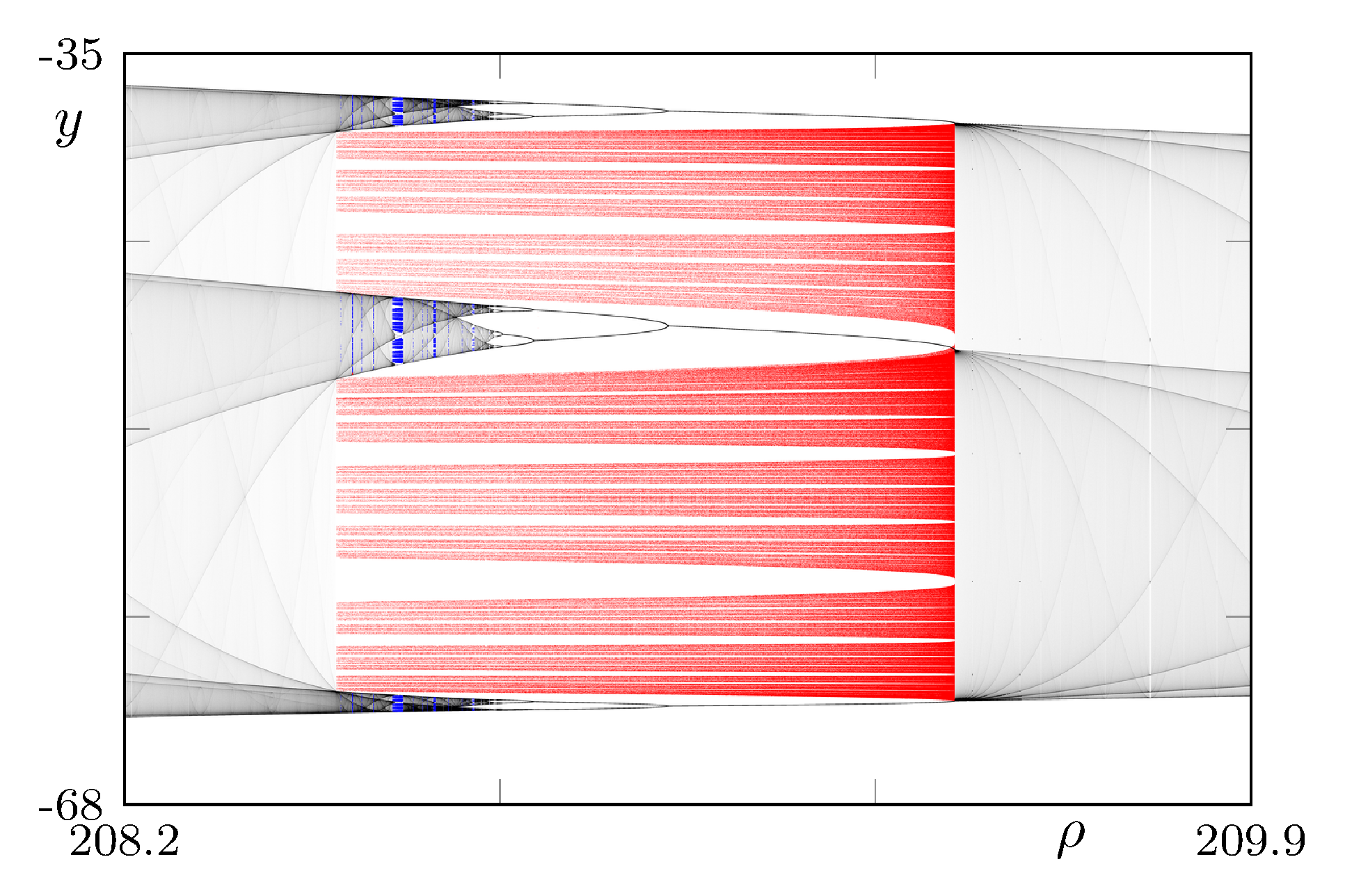}
 \caption{{\bf A periodic window in the graph bifurcation diagram of the Poincar\'e map of the Lorenz system}.
 This figure is placed here for comparison with the very similar Fig.~\ref{fig:p3} for the logistic map. 
 More information about the Lorenz system and its Poincar\'e map is given in the text.
 This window runs from $\rhoo\simeq208.520$ to $\rhoo\simeq209.453$. 
 There is a rectangle in the right side of Fig.~\ref{fig:lfull}(top) that represents the area shown here. 
 In that figure one can see that the red Cantor set and the attractor have components for $y$ outside of the range shown here and that there is another attractor. 
 Several nodes that are unstable Cantor sets are shown in red and blue. 
}
 \label{fig:lcr}
\end{figure}

{\jim We will call a parameter value $\mu_0$ a {\bf cascade value} if there is an infinite cascade of period-doublings limiting $\mu_0$. Fig.~\ref{fig:lmbd} shows a bifurcation diagram where windows are scattered throughout the chaotic region.
The figure shows period-3, period-5, and period-6 windows. There is also a blow-up of part of the period-3 window, in which one sees windows within the period-3 window.
Each of the windows within windows would, with further zooming, reveal a further level of windows within windows within windows, and the process continues ad infinitum.
There is an uncountable set of parameters, each of which is the limit of an infinite nested sequence of windows within windows. 
We call such a parameter value an {\bf infinite-nested value}.

Fig.~\ref{fig:full} shows a graph bifurcation diagram, the same bifurcation diagram with the addition of green points and red points. 
The green points are repelling periodic orbits.
The red points are in repelling chaotic sets.

Figures~\ref{fig:p3} (Logistic map)  and~\ref{fig:lcr} (Lorenz return map) are almost identical, except for a reverse in the direction of the parameter. 
Each shows not only the red chaotic repellors but also a window within a window with blue points that are on a node of an additional chaotic repellor. 
For each window within a window within a window, we expect to see a third chaotic repellor. 


We will argue that, for each cascade value and each infinite nesting value, the graph contains an infinite tower. 
The infinite collection of nodes for such values maybe expected to be a combination of nodes that are chaotic repellors or repelling periodic orbits. 
We summarize these ideas as a conjecture.
}

\smallskip
{\bf Tower Conjecture.} 
\begin{enumerate}
    \item Infinite towers occur within the graphs of chaotic dynamical systems in any dimension {\jim that depend generically on some parameter}.
    \item {\jim More specifically, for generic chaotic dynamical systems depending on a parameter, there would be a countable number of cascade values and an uncountable number of infinite-nested values, each of which has an infinite tower. 
    Furthermore, there is a stable node that is neither periodic nor chaotic, and can be referred to as ``almost-periodic''.
    Such a node can be said to be at the bottom of the infinite tower.
    }
\end{enumerate}
\smallskip

In other words, many chaotic processes have a much more complicated structure than theoreticians previously expected.

{\jim 
The towers described above are not whole story. Sheldon Newhouse proved that chaotic systems can have infinitely many attractors~\cite{New74,Rob83}. 
Of course, each attractor would be a node and there would be no edges between these attractor nodes. 
He showed that, for 2-dimensional maps depending on a parameter, if there is a homoclinic tangency for some parameter value, then there would be uncountably many parameters nearby such that, for each of these, there are infinitely many coexisting attractors. 
These attractors can be very difficult to find numerically.
Even one-dimensional maps can have multiple attractors, see Fig.~\ref{fig:S-bimodal}.
}

We present numerical arguments in support of our conjecture.

The article is structured as follows. 

In Sec.~\ref{sec:Lorenz} we discuss our numerical results on the bifurcation diagrams of the Lorenz map, including the graph bifurcation diagram (Fig.~\ref{fig:lcr}). In particular, we plot the attractor together with some of the repelling chain-recurrent sets and argue that there are parameter ranges where the diagram looks exactly as the one of the logistic map. Our Tower Conjecture is a direct consequence of these observations.

Motivated by these results, in Sec.~\ref{sec:lm} we review some fundamental results on the logistic map and describe the most important features of its graph bifurcation diagram.

{\jim
In Sec.~\ref{sec:numerics} we briefly describe the main numerical algorithms we used to produce the pictures of this article. 
}

Finally, in Sec.~\ref{sec:inf} we describe the graphs of some partial differential equations and differential delay equations. 
All the published results we know of describe the graphs of these systems as being finite and hence simpler than the most complicated cases of the logistic map.


\allblack

\section{The Lorenz system has windows within windows \textit{\textbf{ad infinitum}} {\jim and infinite towers}}
\label{sec:Lorenz}

\begin{figure}
 \centering
\includegraphics[width=11cm]{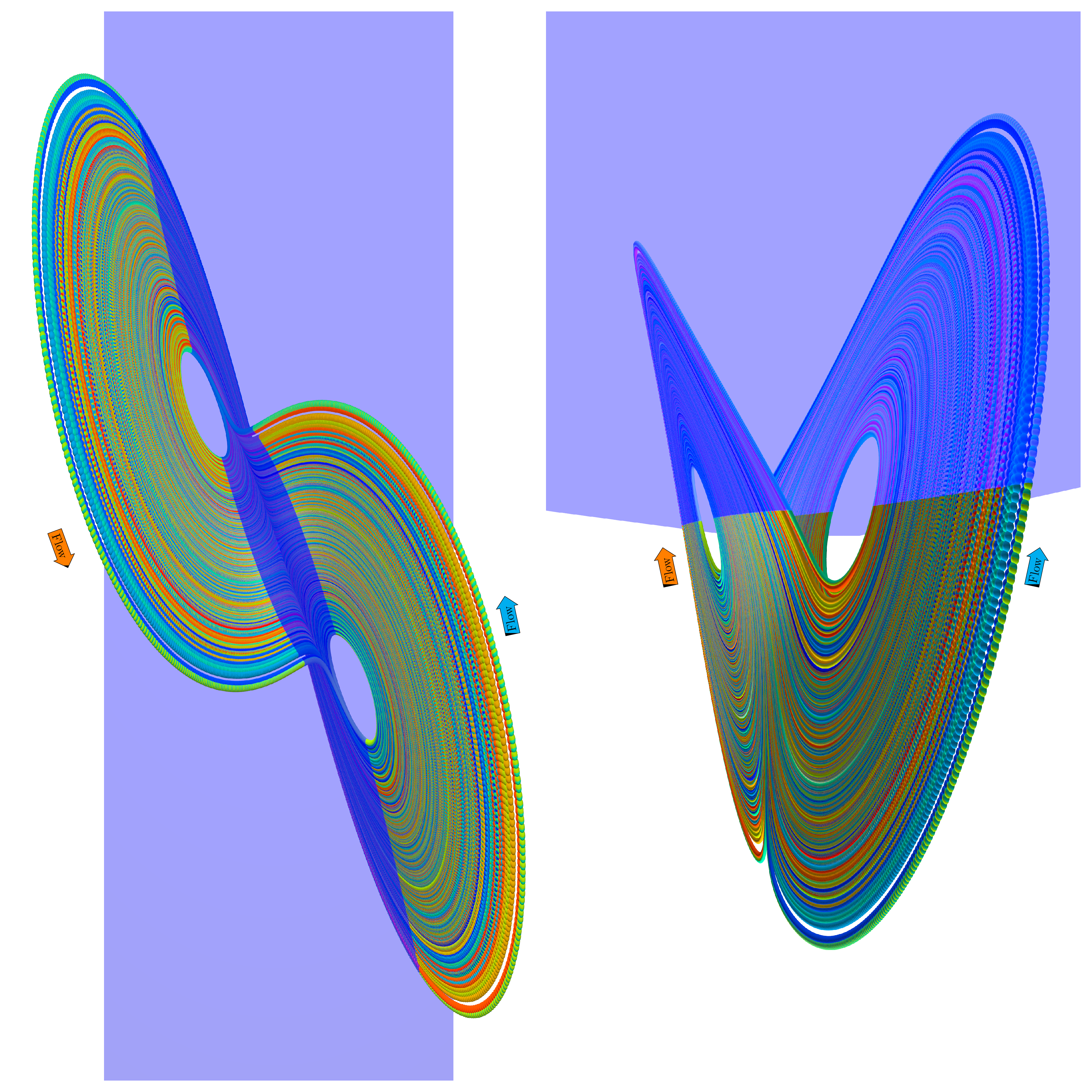}
\centerline{A view from above.\hskip2.5cm A view from below.}
 \caption{{\bf The Lorenz Butterfly}. 
    This picture shows the attractor of the Lorenz system for $r=28$. 
    The color of the trajectory being plotted slowly varies to help visualize the flow.
    If $z$ is thought of as the vertical coordinate, then the left picture is viewed from above and the right one from the side. 
    In the left, blue represents the $z=r-1=27$ horizontal plane; the attractor appears dark blue for points with $z<r-1$. 
    On the right, the attractor is colored with a blue tint when $z>r-1$.
    The Poincar\'e map produces points where the colored attractor meets the blue plane. 
    The arrows indicate the direction of flow. 
 }
 \label{fig:butterfly}
\end{figure} 
\begin{figure}
 \centering
\includegraphics[width=11cm]{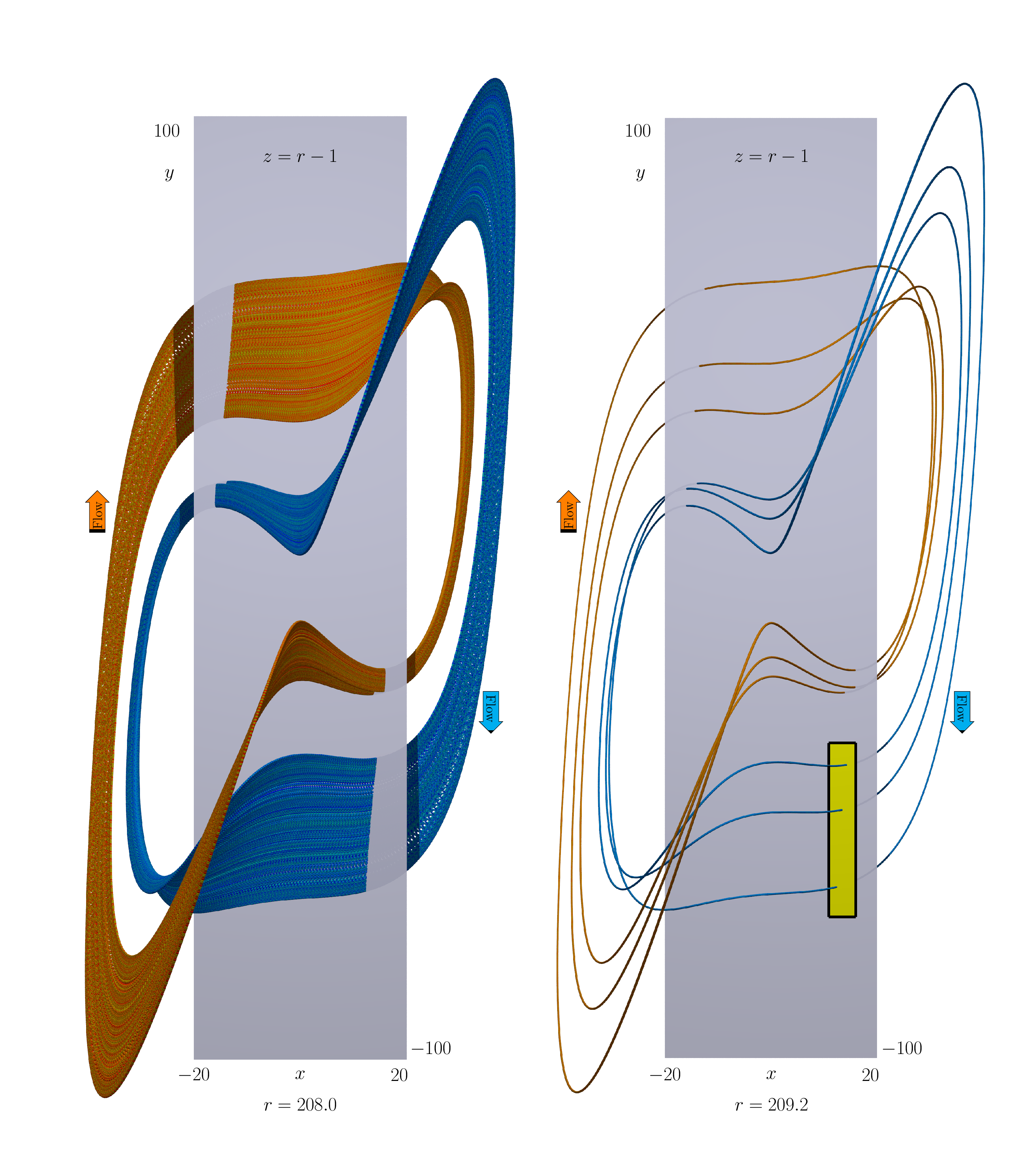}
 \caption{{\bf Attractors of the Lorenz system}. 
    This picture shows the attractors of the Lorenz system for $\rhoo=208$ (left) and $\rhoo=209.2$ (right) together with the rectangle $-20\leq x\leq20$, $-100\leq y\leq100$ in the plane $z=\rhoo-1$. 
    If $z$ is thought of as the vertical coordinate, both pictures are viewed from below.
    The Poincar\'e map for the Lorenz system is built out of the intersections of the Lorenz orbits crossing this rectangle downwards. The yellow rectangle shown for $\rhoo=209.2$ is the one shown (not in scale) in Fig.~\ref{fig:l2dtm} and the three intersections of the blue orbit are at the center of the three little circles shown in that picture. 
 }
 \label{fig:l3d}
\end{figure} 
\begin{figure}
 \centering
 \includegraphics[width=12cm]{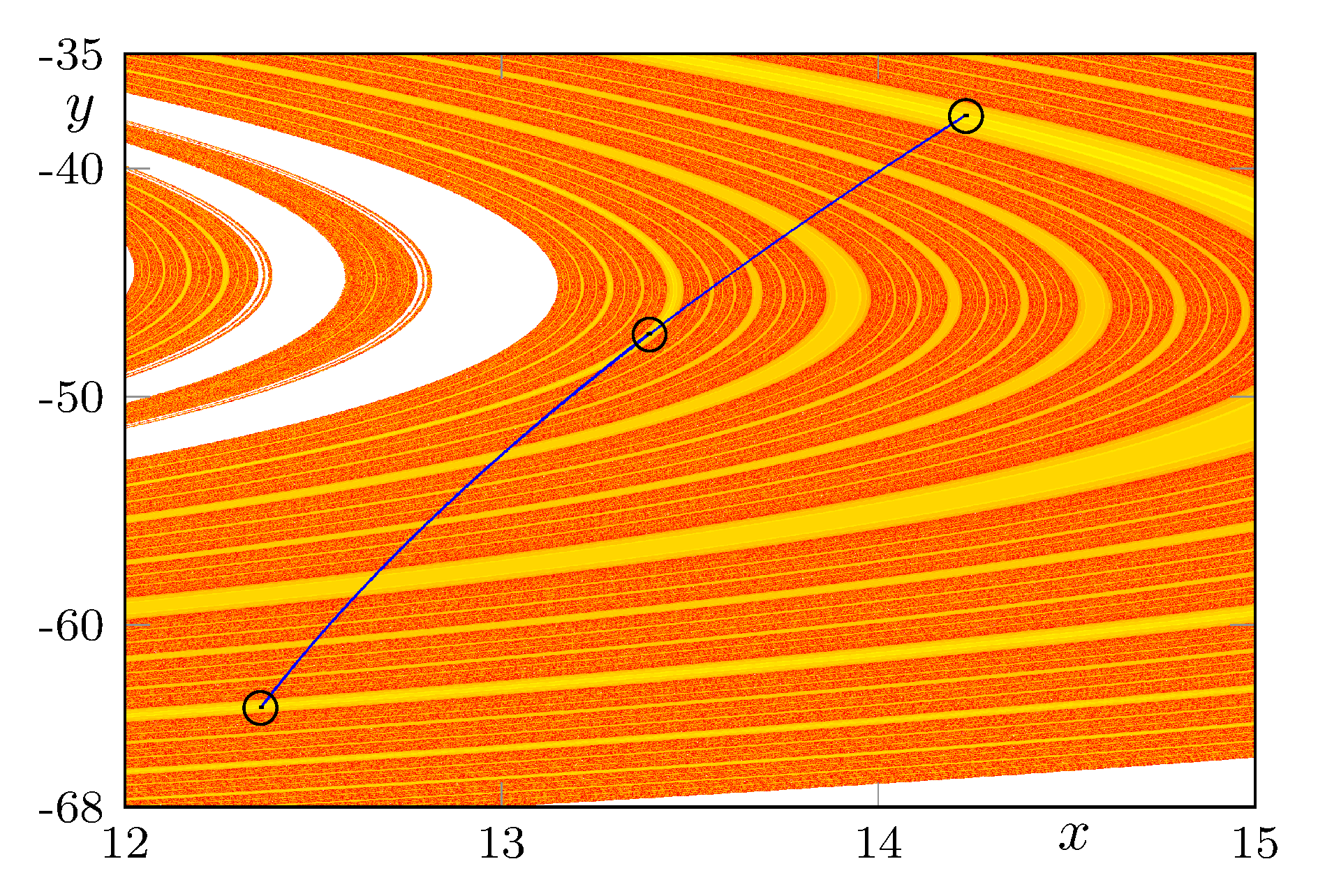}
 \caption{{\bf A small region from the Lorenz system Poincar\'e return map {\BF $P_\rhoo$} for {\BF $\rhoo=209.2$}.} 
 The region shown corresponds to the small yellow region on the right-hand side of Fig.~\ref{fig:l3d}.
 In that figure, a periodic orbit is shown piercing the yellow rectangle in three points.
 Those points are shown here as the centers of three circles. 
 Almost all points in the colored region are in the basin of attraction of the periodic orbit. Yellow indicates rapid convergence to the periodic orbit. Red indicates slow convergence. Red points are close to points that are attracted to the Cantor set on the blue line.
 The blue curve is the unstable manifold of a Cantor set that lies within it.
 Points in the white region are attracted to the other off-screen attractor. 
 The blue curve includes a chain-recurrent Cantor set of saddle points and the unstable manifolds of all of the periodic orbits in the Cantor set. 
 }
 \label{fig:l2dtm}
\end{figure} 
In the 1960s, Edward Lorenz introduced and investigated the ODE system 
\beq
\begin{cases}
\label{eq:Lor}
x' &=-\sigma x +\sigma y  \\
y' &=-xz+\rhoo x -y\\
z' &=\phantom{+}xy - \betaa z  
\end{cases},
\eeq
that is now named after him~\cite{Lor63}, for a specific set of parameters: $\sigma=10$, $\rhoo = 28$ and $\betaa=8/3$. 

It was in the attempt to understand the dynamics behind the Lorenz map that Li and Yorke showed that ``period-3 implies chaos''~\cite{LY75}. Later in the same decade, Yorke and Kaplan~\cite{KY79} showed the presence of chaos, in form of a strange repellor, in the Lorenz system already at $\rhoo$ between 13.9 and 24.06, but this chaotic behavior happens only for a measure-zero set of points. At $\rhoo\simeq24.06$ the repellor becomes an attractor and so the chaotic set has a basin. Shilnikov and collaborators~\cite{ABS77} had closely related results two years earlier.
\begin{figure}
 \centering
 \includegraphics[width=11.8cm]{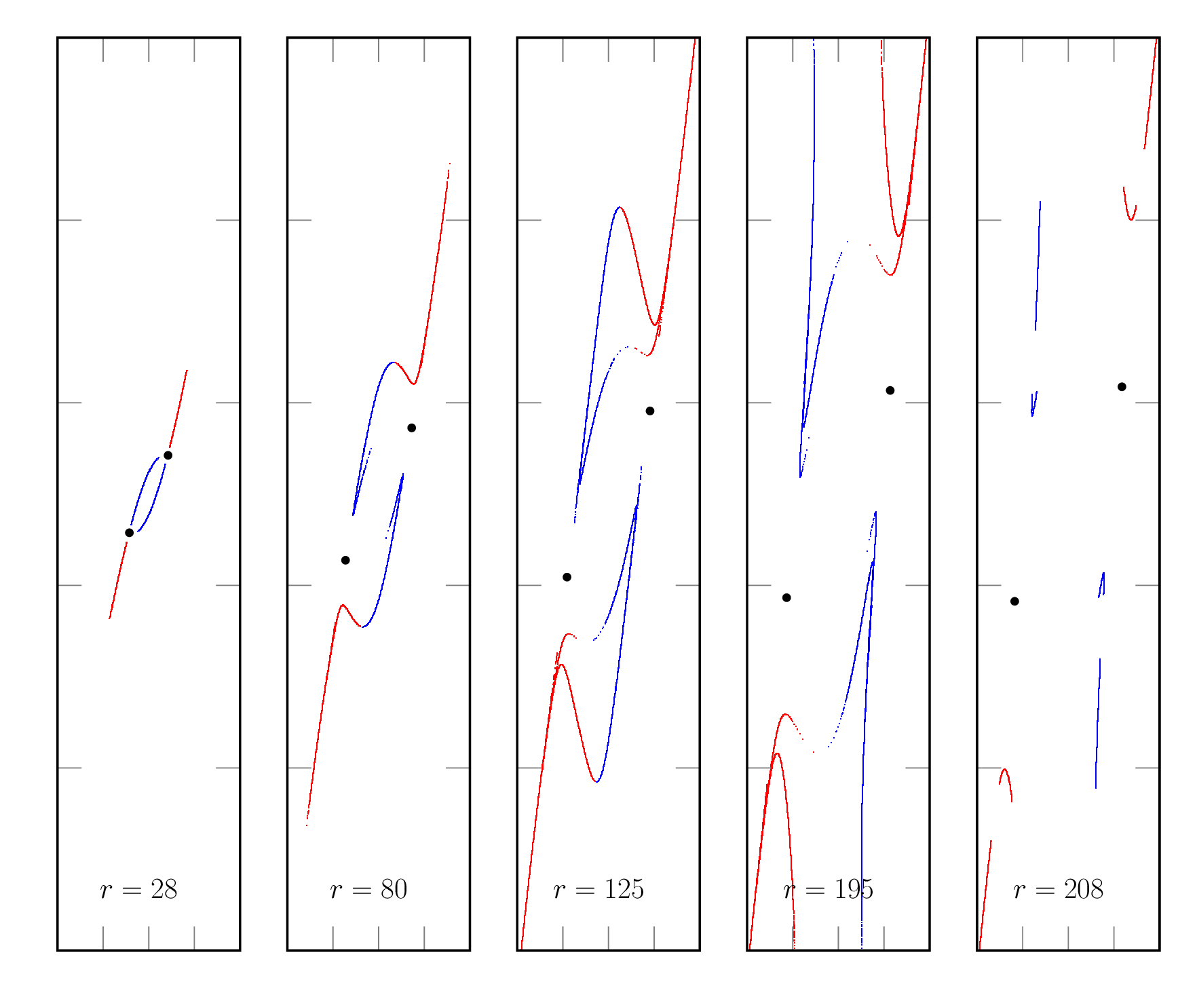}
 \caption{{\bf Attractors of the Poincar\'e Map of the Lorenz system.} 
    The pictures above show the attractor of the Poincar\'e map of the Lorenz system  in the $(x,y)$ plane for five different values of $\rhoo$. 
    Whenever the trajectory hits the $z=r-1$ plane, a point is plotted, in blue if $z$ is decreasing and otherwise in red.
    There are places where the color switches from blue to red, due to the vector field being tangent to the plane. 
    In all pictures $x$ ranges from $-40$ to $40$ and $y$ ranges from $-100$ to $100$. 
    Each panel is the plot of a single trajectory, hence low density regions of the attractor may not be represented or it may only be represented by a few isolated points.
    There are two steady states on this plane (Eq.~\ref{eq:ss}). 
    They are indicated with black dots.
 }
 \label{fig:la}
\end{figure} 

In another work, Ellen Yorke and James Yorke~\cite{YY79} investigated the transition to chaotic dynamics at $\rhoo\simeq24.06$.

Based on these works, Sparrow investigated numerically  the Lorenz system~\cite{Spa82} for a wider range of values of $\rhoo$. His Fig. 5.12 on p.~99 shows intervals of $\rhoo$ values (\ie, windows) where the chaotic attractor is replaced by periodic attractors. He reports that below $\rhoo=30.1$ there are no windows in the bifurcation diagram. 

In 2002, W. Tucker~\cite{Tuc02} proved  rigorously the existence of a strange attractor in the Lorenz System at $r=28$ (see~\cite{Ghy13} for a review of the analytical study of the Lorenz system and its crucial role in the development of chaos theory). 
This is the 14th of the list of  ``mathematical problems for the next millennium'' made by Smale in 1998~\cite{Sma98}. It is noteworthy to mention that the proof of Tucker is computer assisted (see~\cite{Ste00,Via00} for interesting reviews of Tucker's result).
See also rigorous results related to the Lorenz attractor~\cite{Rue76,Ran78,GW79,Wil79}. 
Fig~\ref{fig:butterfly} shows two views of the attractor for $r=28$.

More recently, Kobayashi and Saiki~\cite{KS10,KS14} investigated how periodic windows arise as the parameter $\rhoo$ increases from the Lorenz value and argue that the first windows of the bifurcation diagram are contained in the interval $30\leq\rhoo\leq32$.
\begin{figure}
 \centering
 \includegraphics[width=12.21cm]{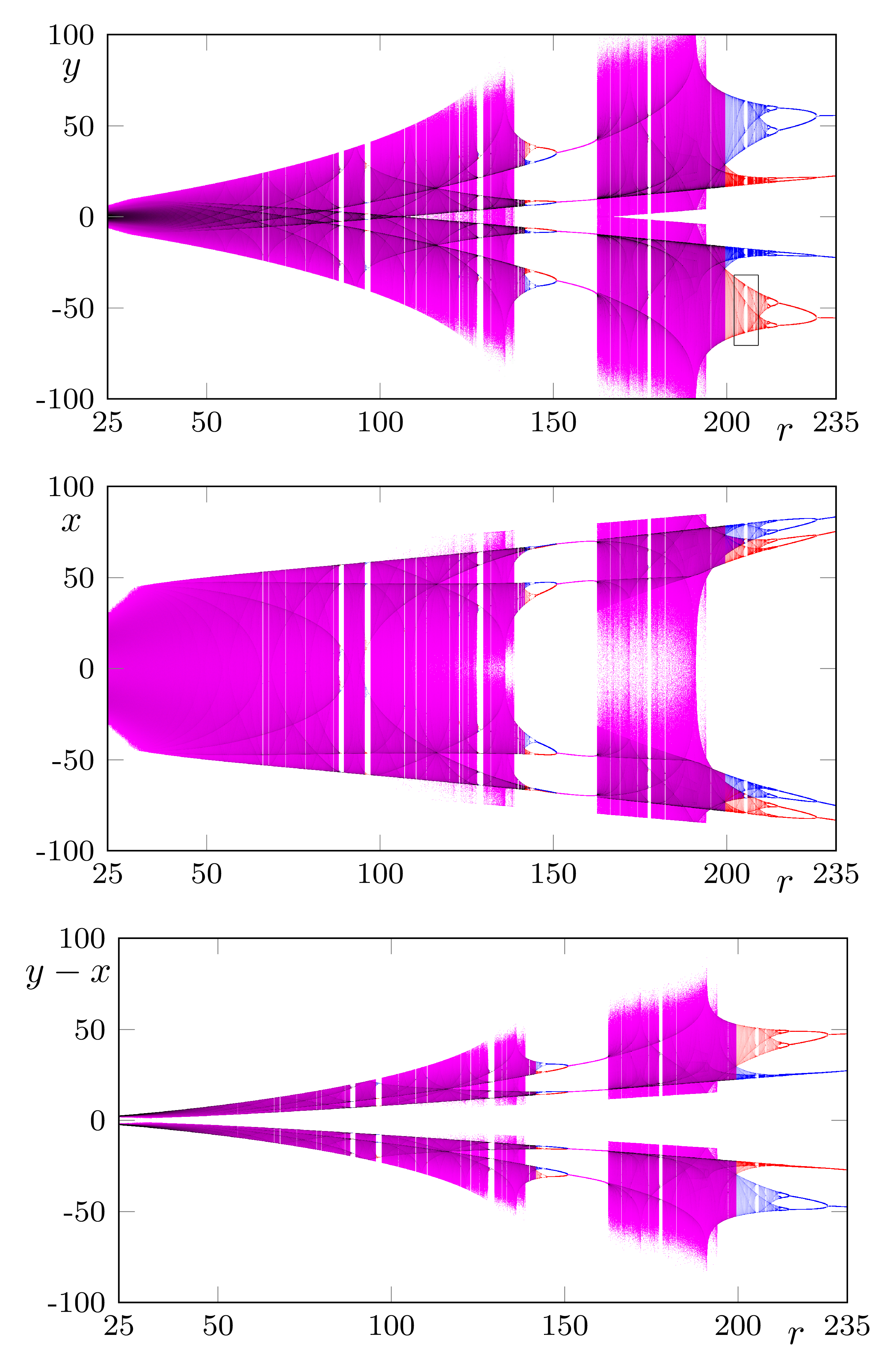}
 \caption{{\bf Bifurcation Diagram}. 
    These are the projections onto the $(r,y)$ plane (Top panel), $(r,x)$ plane (Middle panel) and $(y-x,r)$ plane (Bottom panel) of the bifurcation diagram for the Poincar\'e Return map of the Lorenz equations~(\ref{eq:Lor}) using the plane $\pi_r$ defined by $z=r-1$. 
    A dot is plotted in the $(r,y)$ (resp. $(r,x)$) plane when a trajectory crosses downward past $\pi_r$ through the point $(x,y,r-1)$. 
    The regions where there is speckled white and magenta dots is where the attractor is low density. 
    The Lorenz attractor typically has great variations in density, so extremely long trajectories would be needed to reveal such parts of the attractor. 
    In Fig.~\ref{fig:p4}, we show details of the period-4 window that is centered around $r=150$. 
    The period-3 window shown in Fig.~\ref{fig:lcr} is an enlargement of the black rectangle shown in the top projection.
    For some parameter values, there are two attractors.
    They are shown in red and blue.
    When there is a single attractor, it is shown in magenta.
    }
 \label{fig:lfull}
\end{figure} 

The numerical explorations that we present in this article aim at providing numerical evidence that the structure of the graph bifurcation diagram of the Lorenz system is qualitatively similar to the Logistic map's. 
In particular, the logistic map has parameter values, each of which has infinitely many disjoint unstable invariant sets that are chain-recurrent and form a tower (see Sec.~\ref{sec:lm}). 

\begin{figure}
 \centering
 \includegraphics[width=12cm]{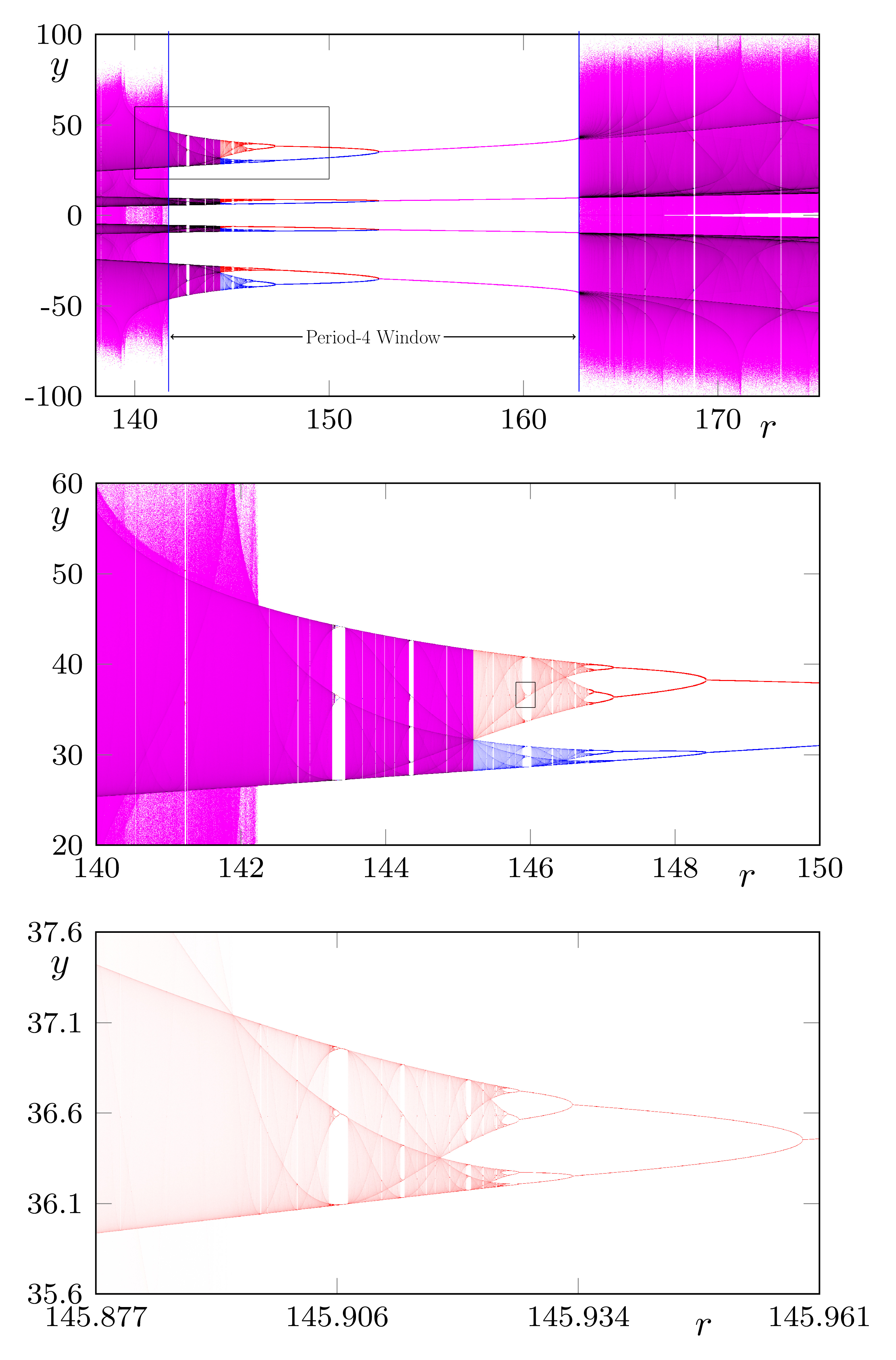}
 \caption{{\bf Bifurcation Diagram}. These $(r,y)$ projections are enlargements of the main period-4 window of the Lorenz system bifurcation diagram. {\bf Top:} We show a zoom of the full window. {\bf Middle:} We show the content of the region enclosed in the black rectangle in the top picture, namely the upper cascade. {\bf Bottom:} We show the middle cascade of the period-3 window of the cascade above, enclosed in a black rectangle in the middle picture.}
 \label{fig:p4}
\end{figure} 

The bifurcation diagram of the Lorenz system is obtained as follows. For every $\rhoo>1$, the Lorenz system has three fixed points: the origin and the twin points 
\beq
\label{eq:ss}
C_\pm=\left(\pm\sqrt{\betaa(\rhoo-1)},\pm\sqrt{\betaa(\rhoo-1)},\rhoo-1\right).
\eeq
The origin is a saddle while $C_\pm$ have a pair of complex eigenvalues.
On the plane $\pi_\rhoo$ defined by $z=\rhoo-1$, integral trajectories passing through points $p$ close to $C_\pm$ will return and cut again the same plane in some other point $q$ and so on. 
As long as the trajectory passes through  $\pi_\rhoo$, it will cut the plane one time directed upwards and the next time directed downwards. 

{\bf Poincar\'e Return Maps.} 
Poincar\'e discussed trajectories that crossed some special plane or line.
When he investigated the planar restricted three-body problem, he found it useful to record only half the crossings, those for which a particular coordinate was increasing. 
He encountered no tangencies to the line. 
We usually take Poincar\'e's approach but, in Fig.~\ref{fig:la}, we record both crossings in two colors, as was done in~\cite{SSY17}. 

We define the Poincar\'e map $P_\rhoo$ at a point $p$ to be the point $q$ at which the trajectory starting from $p$ cuts the plane $\pi_\rhoo$ with $z$ decreasing. 
In Fig.~\ref{fig:l3d} we show two examples of attractors for the Lorenz system, a chaotic one (left) and a periodic one (right). 
The pictures also show, in gray, the rectangle $-20\leq x\leq20$, $-100\leq y\leq100$ in the corresponding planes $\pi_\rhoo$. 
By bifurcation diagram of the Lorenz system we mean the bifurcation diagram of the family of maps $P_\rhoo$.

In Fig.~\ref{fig:lfull} we show a few projections of the bifurcation diagram: on the $(y,\rhoo)$ plane (top), on the $(x,\rhoo)$ plane (middle) and on some intermediate plane (bottom). In particular, the bottom picture suggests that the bifurcation diagram is the union of two disjoint components, one the image of the other. This fact is also suggested by the $(x,y)$ sections of the diagram for several values of $\rhoo$ shown in Fig.~\ref{fig:la}.

The bifurcations pattern of the Lorenz system evolves backwards with respect to the one of the logistic map. 
At $\rhoo=235.0$ the attractors are a pair of period-2 orbits (shown in red and blue), each of which undergoes, as $\rhoo$ decreases, a bifurcation cascade completely analogous to the one of the logistic map. 
The largest window of the diagram (see Fig.~\ref{fig:lfull} and~\ref{fig:p4}), centered at about $\rhoo=150.0$, starts (from the right) with a single period-4 orbit and again contains bifurcation diagrams quite analogous to those of the logistic map. 
Many other windows of smaller different sizes are clearly visible in all three projections.

By zooming on the cascades, the diagram looks more and more like the one of the logistic map. For instance, in Fig.~\ref{fig:p4} we show a full picture of the period-4 window (top), a detail of its upper bifurcation diagram (middle) and a detail of the middle cascade of the period-3 window within it (bottom). Both the red cascade and its sub-cascade look almost identical to the logistic map bifurcation diagram (see Fig.~\ref{fig:lmbd}).

We also investigate the structure of the chain-recurrent set of the diagram. We focus on its largest period-3 window, that is the one contained inside its very first cascade from the right. The range of this window is from  about $208.52$ to $209.453$. In Fig.~\ref{fig:lfull}, it is visible as the largest window within the red and blue cascades at the top and bottom of the diagram. 

In Fig.~\ref{fig:lcr} we show a full size picture of the Lorenz period-3 window. 
The attractor is shown in black/gray while the invariant Cantor sets are shown in red and blue.
Fig.~\ref{fig:p3} shows the analogous picture for the logistic map. 
The structures in the two maps look almost identical.

%
\section{Infinite Towers in the logistic map}
\label{sec:lm}
The logistic map
\beq
\label{eq:lm}
\ell_\mu(x) = \mu x(1-x)
\eeq
is among the simplest continuous maps giving rise to a (highly) non-trivial dynamics. 
In this article
{\bf we focus exclusively on the parameter interval
\BF$\mu \in (1,4)$.} For these values,
$\ell_\mu$ maps $[0,1]$ into itself, 0 is a repelling fixed point and there is exactly an attractor in $(0,1)$.

To simplify the logistic story, {\bf our graphs only include nodes in \BF$(0,1)$.}
In particular, {\bf our graphs omit the fixed point at \BF$x=0$, which is always the top-most node}, and we ignore points outside of $[0,1]$. Their trajectories diverge to $-\infty.$

{\bf Some history.}
In a series of celebrated works starting in 1918, Julia and Fatou gave birth to the study of the dynamics of the quadratic map in the complex plane.
Surprisingly, the study of the quadratic map in the real line began later. The first example we know of is by Chaundy and Phillips~\cite{CP36} in 1936, inspired by early Mathematical Biology works such as~\cite{Bai31}. 

The study of iterations of real quadratic maps reappeared in a few clever abstract articles (``abstract'' in that no applications were mentioned) around 1960 by P.J. Myrberg~\cite{Myr58,Myr62,Myr63}. 
Myrberg discovered the infinite number of period-doubling bifurcations in the logistic map. In the same years, fundamental properties on the existence of cycles for general continuous maps of the real line into itself were discovered by A.N. Sharkovski{\u\i}~\cite{Sha64a} (in Russian; see English translation in~\cite{Sha64b}).

Possibly the first time that $\ell_\mu$ was called the ``logistic map'' was in 1968 in J. Maynard Smith's book ``Mathematical ideas in Biology''~\cite{Smi68}. Smith used it as a toy model for population dynamics, analogous to the one-century old logistic
ordinary differential equation model of Verhulst~\cite{Ver38,VZSC75}. 

In the $1970$'s, many more works on the logistic map appeared in literature, some purely theoretical  (\eg\ Metropolis, Stein and Stein~\cite{MSS73}, Li and Yorke~\cite{LY75}, Hoppensteadt and Hyman~\cite{Hop77}) and some applied (\eg\ May~\cite{May75}, Smale and Williams~\cite{SW76}, May and Oster~\cite{MO76}, Guckenheimer, Oster and Ipaktchi~\cite{GOI77}, Feigenbaum~\cite{Fei78}). 
The celebrated article by R. May~\cite{May76} brought the importance of 1-dimensional dynamics to a broad scientific audience.

The theoretical study of the logistic map evolved in 1980s in the study of families of more general one-dimensional real maps such as, in order of generality, S-unimodal, unimodal and multimodal (\eg~see~\cite{dMvS93,vSt10,Lyu12}).

{\bf The bifurcation diagram.}  The logistic map has exactly one attractor for each parameter value (e.g. see~\cite{dMvS93}, Thm.~4.1, or~\cite{Lyu12}). Bifurcation plots for $\mu$ to the left of the so-called {\bf Myrberg-Feigenbaum } or {\bf Feigenbaum  parameter value} $\mu_F\simeq3.5699$~\cite{Fei78} appeared in several publications in the $1970$'s but, to the best of our knowledge, the first picture of the full bifurcation diagram (Fig.~\ref{fig:lmbd}) appeared first in an article by Grebogi, Ott and Yorke in 1982~\cite{GOY82}.

Usually, bifurcation diagrams show how the attractors change with the parameter, just as in Fig.~\ref{fig:lmbd}. {\jim In this article, however, we also include some graph bifurcation diagram (Figs.~\ref{fig:full}--\ref{fig:lcr})}.

The bifurcation diagram starts with an infinite cascade of period doublings at the values $\mu_0=1$, $\mu_1=3$, $\mu_2=1+\sqrt{6} \simeq3.4495,\dots$, whose speed increases exponentially until the Myrberg-Feigenbaum parameter value $\mu_F\simeq3.5699$. 
This is the border after which there is chaos. There are ``period-doubling'' parameter values $\mu_n$ such that for $\mu_n<\mu<\mu_{n+1}$, the attractor is a periodic cycle of $2^n$ distinct points. Feigenbaum's fundamental discovery was  that the speed of the bifurcation cascade 
\beq
\lim_{n\to\infty}\frac{\mu_n-\mu_{n-1}}{\mu_{n+1}-\mu_n}\simeq4.6692
\eeq
is universal,  {\jim only in the sense that the same limit is obtained for a large class of systems that have a period-doubling cascade. It is found not only in one-dimensional but also in higher-dimensional non-Hamiltonian maps.} 
(Hamiltonian processes, however, yield different numbers).
Sanders and Yorke~\cite{SY13} proved that cascades of period doublings are quite ubiquitous in low-dimensional dissipative systems.  

\subsection{The three kinds of attractors for the logistic map.}

In~\cite{Guc79} Guckenheimer proved that, for every value of $\mu$ in $[0,4]$, the logistic map $\ell_\mu$ (or, to be precise, any S-unimodal map) has exactly one attractor and that this attractor must be precisely of the following three kinds. 

{\bf First kind: a periodic orbit.} 

{\bf Second kind: a finite union of intervals.} 
In this case, the attractor is a collection of intervals $J_1,\dots,J_n$ such that $\ell_\mu(J_i)=J_{i+1}$ except that $\ell_\mu(J_n)=J_{1}$. Furthermore the map is chaotic. 
Most often, there is a single interval $J_1$ and $\ell_\mu(J_1)=J_{1}$.
In particular, for most $\mu$ between $\mu\INT$ and 4 (see Fig.~\ref{fig:lmbd} for $\mu\INT$), the attractor is an interval and the dynamics on it is chaotic. 
However, there are windows, \ie, intervals in parameter space, where the attractor is not an interval.
Now, also in Fig.~\ref{fig:lmbd}, each window has a bifurcation diagram that is tiny but extremely similar to the entire diagram (see Subsec.~\ref{Windows}). Such windows occur not only after $\mu\INT$ but more generally after $\mu_{FM}$.

\begin{figure}
 \centering
 \includegraphics[width=8cm]{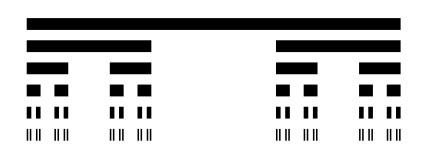}\ \caption{{\bf Example of Cantor set.} First steps in the construction of the standard Cantor subset of $[0,1]$, obtained by eliminating recursively the central third part of all segments. 
 The invariant Cantor sets of the logistic map are much less regular than the example shown here. 
 }
 \label{fig:cantor}
 \end{figure}

{\bf Third kind: a Cantor set attractor.}
We call this attractor {\bf almost periodic}~\cite{Mil85}, also sometimes called ``odometer''. This kind of attractor is a Cantor set and it is not chaotic.
It occurs precisely when the graph  has infinitely many nodes. 
Each node, other than the attractor, is either an unstable periodic orbit or a repelling chaotic Cantor set.

For each point $x_0$ in the attracting Cantor set and each $\varepsilon>0$, there is a periodic point $x_{\varepsilon}$ such that the $n^{\text{th}}$ iterate of the map on $x_0$ and the $n^{\text{th}}$ iterate of the map on $x_{\varepsilon}$ stay within $\varepsilon$ of each other for all time. 
At $\mu_{FM}$, every period orbit has period $2^n$ for some $n$ and none of them belongs to the Cantor set. 
They converge to the Cantor set as $n\to\infty$.

Recall that any Cantor subset $C$ in an interval $I$ is an uncountable set that contains no intervals. 
Also, it has no isolated points in the sense that each point in the set is a limit of other points in the set. 

\medskip
In all three cases, ``almost every'' $x\in(0,1)$ belongs to the basin of attraction.
By almost every we mean that the points that are not attracted can be covered by a finite or countably infinite collection of intervals with arbitrarily small total length.

Notice that, in case of an attracting Cantor set, the basin has empty interior. Each open neighborhood of such an attractor contains infinitely many nodes of the graph.

We can write the parameter space as $(1,4]={\cal A}_P\cup\AC\cup\ALP$, where the union is disjoint, ${\cal A}_P$ is the set of parameters for which the attractor is a periodic orbit, $\AC$ is the set of parameters for which the attractor is chaotic and $\ALP$ the set of those for which it is a Cantor set  (see~\cite{JR80}).

The set ${\cal A}_P$ is open, which simply follows from the stability of attracting cycles under small perturbations, and dense, which instead requires heavy machinery and was proved rigorously only in 1997, independently by Lyubich~\cite{Lyu97} and Graczyk and Swiatek~\cite{GS97}. Heuristically, the density of this set follows from the fact that it is to be expected that, for almost all chaotic parameters, the orbit of the critical point $c$ will be dense in the attractor and so, in particular, its orbit will get arbitrarily close to itself. 
This way, arbitrary small changes in $\mu$ should be able to make the orbit of $c$ become periodic~\cite{Ott02}, and a periodic orbit containing the critical point is always super-attracting and lies within a regular window.

Since $\AP$ is dense and it contains $(1,\mu_{FM})$, its complement $\AC\cup\ALP$ is a Cantor subset of $[\mu_{FM},4]$. Jakobson~\cite{Jak81} proved in 1981 that $\AC$ has positive measure while it was proved only in 2002 by Lyubich~\cite{Lyu02} that $\ALP$ has measure zero.

Notice that all results above hold not just for the logistic map but also for every non-trivial real analytical family and any {\em generic} smooth family of unimodal maps~\cite{dM04,SvS14}.
However, it is known that there are non-generic smooth families showing ``robust chaos''~\cite{BGY98,AA01,vS10}, namely without windows.

{\bf The edges of the \gr graph.} 
Our approach to the logistic map aims at finding the nodes, the chain-recurrent sets, and the edges of the graph.

Most of the traditional literature on logistic maps look at the  non-wandering sets.
One non-wandering set can contain many nodes.
Most of the ideas are similar but one must be careful.
The non-wandering set of unimodal maps was first described by Jonker and Rand in 1980~\cite{JR80} (see also~\cite{vS81,HW84,BL91,dMvS93,Blo95,SKSF97}). 

Furthermore, no one seems to have examined the edges of the graph; therefore we do in~\cite{DLY20} where, as mentioned above, 
we prove the following.

{\bf 
For every {\BF$\mu\in (1,4]$}, the graph is a tower.
In particular, there is an edge between every pair of nodes.}

To illustrate this idea, we now argue that the top node always consists of the point 0, except for $\mu=4$. For any point $x_0>0$ close enough to 0, there is a backward trajectory $x_n$ with $n<0$ that converges to $0$ as $n\to-\infty$. For $\mu\in(1,3]$, there is a unique attractor, which is the non-zero fixed point $p$ of $\ell_\mu$, and 
the trajectory $x_n$ converges to $p$ as $n\to+\infty.$
That means there is a edge from 0, which is a node, to the attractor node. Of course, for $\mu = 1$, the attractor is 0.

For $\mu\in[3,4]$, choose any $x_0>0$ near 0. Write $J$ for the interval $[0,x_0]$.
When we apply the map $\ell_\mu$ to $J$, we obtain a longer interval and, as we repeatedly apply the map, we eventually obtain an interval that includes $[0,\frac{1}{2}]$.
Notice that $\frac{1}{2}$ is critical point of the logistic map.
Hence with one more iterate, the image interval is 
$[0, \ell_\mu(\frac{1}{2})].$
For any point $p$ in any node, there is a point $\hat x_0$ near 0 such that there is a $N$ for which $\ell_\mu^N(\hat x_0) = p$. 
Since each node is an invariant set, the forward limit set of $\hat x_0$ is in the node.  
As discussed above, there is a backward trajectory from $\hat x_0$ that limits on $0$ as $n\to-\infty$. 
Hence the graph has an edge going from $0$ to the node containing $p$. 
That is true for every $p$ in every node.
Hence, for each node other than $0$,
there is an edge from $0$ to that node.
For $\mu=4$ there is a single node, that is the whole interval $[0,1]$.

{\bf Because of this fact, to simplify the pictures of graphs, in Figs.~\ref{fig:full},\ref{fig:p3} we do not include the ``zero node'', the node that consists of $0$, which is always on top of each tower.}
\subsection{Windows in the bifurcation diagram}
\label{Windows}
{\bf A period-\BF$ k$ window (of parameter values).} Figure~\ref{fig:p3} shows a ``period-3 window'', an interval of parameters in which there are three intervals $J_i$, $i=1,2,3$, in $x$ space which are permuted by the map. 
Each of the intervals $J_i$ changes continuously, starting at the parameter value  $\mu=1+\sqrt{8}\simeq3.8284$, where the period-3 orbit appears in a saddle-node bifurcation. 
The window ends at the parameter $\mu$ at which the attractor fills the interval, namely $\mu\simeq3.8568$.

Each saddle-node bifurcation of a  period$-k$ orbit begins an analogous {\bf  window} or {\bf period\BF$-k$ window}
with a final $\mu$ at which the attractor fills the intervals $J_i.$

Between the first Feigenbaum parameter value and $\mu=4$, there are infinitely many windows and every $\mu$ in that range is either in a window or arbitrarily close to one. In Fig.~\ref{fig:full} several are visible. 
The biggest is the period-3 window, which is also shown in Fig.~\ref{fig:p3}.

For each parameter value inside a period-$k$ window, there are $k$ intervals $J_j$ mentioned above and there is a chain-recurrent set $C$ of points whose forward trajectories do not fall into any of the the $J_j$. 
These sets are shown in red for the larger windows in Fig.~\ref{fig:full}. The set $C$ is always a Cantor set.

{\bf Windows within windows within windows \ldots. } In Fig.~\ref{fig:p3} we see the period-3 window. 
The bifurcation diagram of the attractor is plotted in black and gray. 
It consists of three pieces that each look like the entire bifurcation diagram. 
Each piece lies within one of the $J_j$ for each $\mu$. 
As such, there are windows within this bifurcation diagram, infinitely many windows within the primary window. 
Each windows has secondary windows within it.

For $\mu$ that has a period$-k_1$ window, there are $k_1$
intervals inside which the attractor lies. We denote them by $J_1,\dots,J_{k_1}$. There is a chain-recurrent Cantor set $C^1$ of points that do not fall into them.
If $\mu$ has a period$-k_2$ window within the window, then $k_1$ divides $k_2$ and there are $k_2$ intervals $J'_1,\dots,J'_{k_2}$ that are a small-scale version of the $J_j$ above.

The Cantor set $C^2$ for this window lies outside the union $\cup J'_j$ but inside the union  $\cup J_j$; see Fig.~\ref{fig:p3}. 
Such a $C^2$ Cantor set is shown in blue where the primary window has $k_1=3$ and the secondary window has $k_2= 3\times3 =9.$

The graph has a node $0$ on top of $C^1$ on top of $C^2$ followed by possible more Cantor set from further windows within windows and finally some attractor at the bottom.

{\jim
For each parameter that has windows within windows {\em ad infinitum}, the graph is an infinite tower. 
}

\allblack
\bigskip\noindent{\bf Building blocks of towers.}
If $\mu$ does not belong to any window, then the only chain-recurrent sets are the left endpoint 0 and the attractor.

Just after the start of a window, the attractor is a periodic orbit with some period $k$ (e.g. see T2 in Fig.~\ref{fig:full}). 
As $\mu$ increases, the periodic orbit goes through a bifurcation process identical to the one at the left of $\mu_{FM}$ (e.g. see T3--T4 in Fig.~\ref{fig:full} and T2--T5 in Fig.~\ref{fig:p3}). 
This reflects in the graph in the following way. 
Each subwindow corresponds, in the graph, to a Cantor set node. 
Each Cantor node may be immediately followed by $s$ nodes that are unstable periodic orbits. The periods of these orbits are, in the following order, $k$, $2k$, $2^2k$, up to $2^{s-1}k$. 
After those nodes, there will be either the attractor or another repelling Cantor set node. 

Figures~\ref{fig:full} and~\ref{fig:p3} show examples of towers.
In summary, a graph can contain any number of Cantor set nodes, including none and infinitely many. 
Between two consecutive Cantor set nodes, there can be any finite number of repelling periodic orbit nodes, including no such orbits.
In particular, there can be infinitely many nodes and any combination of Cantor sets and repelling periodic orbits is possible.
All possible finite or infinite patterns (including patterns with only saddles and patterns with only Cantor sets) occur for appropriately chosen parameter values.
In addition, (1) there is an attractor, the bottom-most node; (2) for the logistic, the top node is the repelling fixed point 0 for all $\mu<4$.
{\jim
\section{Numerical Algorithms}
\label{sec:numerics}

We describe here briefly the algorithms we use to generate the pictures of the graph bifurcation diagrams. In all figures, we discretize the space coordinate and the parameter coordinate. We call an elementary cell of this discretization a {\bf pixel}.
When we say we are plotting a Cantor set, the goal is to plot a pixel if that pixel contains a point of the Cantor set. Such statements apply to everything plotted.

{\bf  Shading pixels for attractors.} 
For the pictures of the attractors, in gray in Figs.~\ref{fig:lmbd}--\ref{fig:lcr} and in colors in Figs.~\ref{fig:lfull},\ref{fig:p4},
for each discretized parameter value ($\mu$ in case of the logistic map, $r$ in the Lorenz case)  we iterate the map for a generic initial point and count the number of times the trajectory enters each pixel and then a gray scale is chosen for each pixel in proportion to the pixel's count. We mention this because it has been the common practice in journal figures to color each pixel black if the count is positive and white otherwise.
The gray scale of the picture shows a glimpse of the relative invariant density: the darker the dots, the longer a generic point spends time nearby that pixel.

{\bf Repelling Cantor set for the Logistic map.} First the attractor pixels are identified for each parameter value to be plotted.
The pictures of the Cantor set repellors (in red or blue in Figs.~\ref{fig:full}--\ref{fig:lcr}) are obtained as follows. For each pixel not belonging to an attractor, write the $x$ coordinates of the pixel as $J=[x_*,x^*]$. We examine the interval $J_n$ that runs from $f^n(x_*)$ to $f^n(x^*)$ and plot the pixel if $J_n$ intersects $J$ for some $n>0$.

{\bf Repelling Cantor set for the Lorenz return map.} The Lorenz case (Fig.~\ref{fig:lcr}) needs some more explanation because the return map has a 2-dimensional phase space. 
Fig.~10 shows a region of the phase space of the Lorenz return map.
For a given value of the parameter $r=209.2$ the picture shows a region in the $(x,y)$ plane containing the three points of the period-3 orbit associated to that period-3 window.
The trajectory of some of these points leave the region and do not return. Such points are white.

The picture has a relative global attractor to which all points tend if they remain in the region. 
We discretize the region with a grid 3000 wide by 2000 high. We apply 20 iterates of the return map to each of these grid points and the result is the thin arc plotted in blue. 
Notice the three circles represent three points of a period 6 orbit, where the other three points are outside the plotted region. 
If we think of this blue curve as exactly an arc, each point on the arc has a unique $y$
value. 
The blue arc contains the 3 attracting points and a repelling Cantor set. 
Hence we can apply our methods from the Logistic map to identify points on the Cantor set (to the precision of the grid).

}
\section{Dynamical systems with infinite dimensional phase space can have graphs simpler than the logistic's}\label{ddes-pdes}
\label{sec:inf}

Here we report on examples of graphs that have been determined for differential delay equations and partial differential equations. These examples do not exhibit chaos in the regimes where the graphs have been determined, but the reader should expect great complexity in other examples that have chaos.

\allblack
{\bf Example 1: Delay-Differential Equations.} In 1986, J. Mallet-Paret~\cite{MP88} (see also~\cite{KY75,KY77,McM96,HMW06,HL13}) showed that the graph approach can be applied also to the infinite dimensional dynamical system associated to first-order scalar delay-differential equations of the form
\begin{linenomath}
\begin{equation}
 \dot x(t) = f(x(t),x(t-1)),
 \label{eq:dd}
\end{equation}
\end{linenomath}
with the initial condition $x(t)=\phi(t)$ on $[-1,0]$ for some continuous function $\phi$.
For instance, the celebrated Wright's equation 
\begin{linenomath}
\begin{equation*}
\dot x(t)=-\alpha\,x(t-1)(1-x(t)),\,\alpha>0,
\end{equation*}
\end{linenomath}
modeling population dynamics, is of this form, as well as equations of the form $\dot x(t)=-\alpha x(t)-g(x(t-1))$ that arise in various applications in biology, physiology and optics~\cite{MP88}. 

The \gr graphs for this type of systems are towers with an arbitrary {\em finite} number of levels. 
From the point of view of the dynamics outside of the nodes, therefore, these systems are simpler than some logistic map. 
Below we describe in some detail the nodes of these graphs. 

Denote by $x_\phi(t)$ the solution, defined up to some $T>0$, of~(\ref{eq:dd}). This defines a dynamical system $\Phi^t$, for $t\geq0$, on the space of continuous functions $C^0([-1,0])$,
via 
\begin{linenomath}
\begin{equation*}
\left(\Phi^t\phi\right)(\tau) = x_\phi(t+\tau), -1\leq \tau\leq0.
\end{equation*}
\end{linenomath}
We assume the following properties, satisfied in many applications: 
\begin{enumerate}
\item $f$ is smooth; 
\item $y\,f(x,y)>0$ for all $y\neq0$; 
\item $f_x(0,0)>0$; 
\item $f_x(0,0)+f_y(0,0)>0$; 
\item the image under $\Phi^1$ of any bounded ball is bounded;
\item $\sup_{\phi\in C^0([-1,0]),t\in\mathbb R}\|\Phi^t\phi\|<\infty$.
\end{enumerate}

Under these conditions, the solutions of Eq.~(\ref{eq:dd}) oscillate about zero, the map $\Phi^1$ is compact and dissipative~\cite{HL72} and the flow has a maximal compact attractor ${\cal S}$~\cite{BLS70} equal to the set of all initial conditions $\phi\in C^0([-1,0])$ such that $x_\phi$ is global and bounded.

The decomposition found by Mallet-Paret is relative to the dynamics of the restriction of $\Phi$ to $\cS$.
The unstable nodes of $\Phi$ in $\cS$ are rapidly oscillating unstable periodic orbits. The more rapidly oscillating nodes are above the more slowly oscillating nodes. 
Mallet-Paret has a precise definition of rapidly oscillating, based on the number of zeros the trajectory has on every interval $[t,t+1]$.
He was able to prove that there are orbits joining every node with all nodes that are oscillating more slowly.

{\bf Example 2: Parabolic partial differential equations (PDEs).}
The setting of nonlinear parabolic PDEs proved to be an unexpectedly rich source of dynamical system graphs~\cite{HMW06,Lap19}.

Let $X$ be a closed segment and denote by $H^1(X)$ the Sobolev space of square-summable functions on $X$ with a weak derivative and whose first weak derivative is also square-summable. 
The set $H^1_0(X)\subset H^1(X)$ is closure, in $H^1(X)$, of the set of smooth functions that are zero in some neighborhood of the end-points of $X$. 

We begin with the Chafee-Infante PDE on $X=[0,\pi]$, namely
\begin{equation}
 \begin{cases}
  u_t=u_{xx}+\lambda (1-u^2)u,&\\
  u(t,0)=u(t,\pi)=0\hbox{ for all }\;t\geq0,&\\
  u(0,x)=u_0(x)\in H_0^1(X),&\\
 \end{cases}
\end{equation}
with $\lambda\geq0$. 
We denote by $\Phi^t_\lambda:H_0^1(X)\to H_0^1(X)$ the semiflow of the Chafee-Infante PDE. Then the map $\Phi^{1}_\lambda$ is a $C^2$ (infinite-dimensional) Morse-Smale map for every $\lambda$ which is not the square of an integer~\cite{CI74,HMW06}. In particular, each node of its graph is a fixed point, and the dynamics elsewhere is gradient-like. One of the key observations, by Henry~\cite{Hen85}, leading to the construction of a Lyapunov function for the system, is that the number of components of the set
$
\{x\in X:\,u(t,x)\neq0\}
$
is a monotonously decreasing function of time.

The structure of the \gr graph in this case is more interesting and shows (see Fig.~\ref{fig:co} (right)) the following {\bf bistable} behavior: for $(n-1)^2<\lambda<n^2$, $\Phi^t_\lambda$ has $2n-3$ repelling fixed points $N_1$, $N^\pm_j$, $2\leq j\leq n-1$ and two attracting ones $N^\pm_n$. 
The node $N_1$ has edges to $N^\pm_2$. All $N^\pm_{j}$, $j<n$, have edges towards both $N^\pm_{j+1}$, as shown in the figure.
Note that, in this case, no other edges arise due to a {\em blocking connections} principle that holds for these systems (see~\cite{FR96} for a thorough discussion and examples). 
In particular, in this case the graph is not a tower.


These results do not depend strictly on the analytical form of $(1-u^2)u$ but rather hold for all $C^2$ functions $f(u)$ with a similar shape (see~\cite{CI74,HMW06,Lap19}) and hold for several other important PDEs such as FitzHugh-Nagumo equation and the Cahn-Hilliard equation (see~\cite{Lap19}). 
They were also further generalized by Chen and Polacik~\cite{CP95} to the time-periodic non-autonomous version of the Chafee-Infante equation.

Fiedler and Rocha~\cite{FR96}, finally, further generalized the PDEs above to the autonomous semilinear variation
\begin{linenomath}
\begin{equation*}
u_t=a(x)u_{xx}+f(x,u,u_x)
\end{equation*}
\end{linenomath}
with Neumann boundary conditions $u_x(t,0)=u_x(t,1)=0$.
Denote by $H^2([0,1])$ the Sobolev space of all functions that are square-summable together with their first and second derivatives. 
This PDE form arises in many applications such as population dynamics, astrophysics and material sciences (see~\cite{FR00} for references).

{\jim
\section*{Acknowledgments} This work was partially funded by NSF grant \#1832126. The authors are  grateful to Todd Drumm and Michael Jakobson for helpful conversations on the paper's topic.  

\section*{Conflict of Interest} The authors declare that they have no conflict of interest.
}
\bibliography{refs}
\end{document}